\documentclass[12pt]{article}

\usepackage{amssymb}
\usepackage[dvips]{graphicx}

\newcommand {\beq}{\begin{equation}}
\newcommand {\eeq}{\end{equation}}
\newcommand {\beqa}{\begin{eqnarray}}
\newcommand {\eeqa}{\end{eqnarray}}
\newcommand {\n}{\nonumber \\}
\newcommand {\tr}{\mbox{tr}}

\def\pa{\partial}
\renewcommand{\theequation}{\thesection.\arabic{equation}}

\begin{document}
\setlength{\oddsidemargin}{0cm}
\setlength{\baselineskip}{7mm}

\begin{titlepage}
\renewcommand{\thefootnote}{\fnsymbol{footnote}}
\begin{normalsize}
\begin{flushright}
\begin{tabular}{l}
OU-HET 510\\
December 2004
\end{tabular}
\end{flushright}
  \end{normalsize}

~~\\

\vspace*{0cm}
   \begin{bf}
    \begin{Large}
       \begin{center}
         {ZZ brane amplitudes from matrix models}
       \end{center}
    \end{Large}
   \end{bf} 
\vspace{1cm}

\begin{center}
           Akira S{\sc ato}\footnote
            {
e-mail address : 
akira@het.phys.sci.osaka-u.ac.jp}
            {\sc and}
           Asato T{\sc suchiya}\footnote
           {
e-mail address : tsuchiya@phys.sci.osaka-u.ac.jp}\\
      \vspace{1cm}
       
       {\it Department of Physics, Graduate School of  Science}\\
       {\it Osaka University, Toyonaka, Osaka 560-0043, Japan}
\end{center}

\vspace{2cm}

\begin{abstract}
\noindent
We study instanton contribution to the partition function of 
the one matrix model in the $k$-th multicritical region which corresponds to
the $(2,2k-1)$ minimal model coupled to Liouville theory.
The instantons in the one matrix model are given by local extrema of
the effective potential for a matrix eigenvalue and identified with the
ZZ branes in Liouville theory.
We show that the 2-instanton contribution
in the partition function is universal 
as well as the 1-instanton contribution and  that
the connected part of the 2-instanton
contribution reproduces the annulus amplitudes between the ZZ branes 
in Liouville theory.
Our result serves as another nontrivial check on the correspondence 
between the instantons in the one matrix model and  the
ZZ branes in Liouville theory, 
and also suggests that
the expansion of the partition function in terms of the instanton numbers
are universal and gives systematically
ZZ brane amplitudes in Liouville theory.
\end{abstract}
\vfill
\end{titlepage}
\vfil\eject

\setcounter{footnote}{0}

%%%%%%%%%%%%%%%%%%%%%%%%%%%%%%%%%%%%%%%%%%%%%%%%%%%%%%%%%%%%%%%%%
\section{Introduction}
%%%%%%%%%%%%%%%%%%%%%%%%%%%%%%%%%%%%%%%%%%%%%%%%%%%%%%%%%%%%%%%%%
\setcounter{equation}{0}
\renewcommand{\thefootnote}{\arabic{footnote}}
The nonperturbative study of noncritical strings in terms of the matrix models
\cite{BK,DS,GM} led to an important suggestion \cite{Shenker}
that string theory in general possesses
the nonperturbative effect that behaves as $\sim e^{-\frac{1}{g_s}}$.
It was pointed out in \cite{Polchinski}
and is now widely believed that this effect
is attributed to D-branes.
Indeed, the discovery of D-branes in Liouville theory which are called
the FZZT branes \cite{FZZ,Teschner,PT} and the ZZ branes \cite{ZZ}  
triggered recent progress
in noncritical strings and the matrix models, in which
the origin of the nonperturbative effect 
$e^{-\frac{1}{g_s}}$ given by the string equations of the matrix models
\cite{Shenker,David,EZ} was identified with the ZZ branes \cite{MV}-\cite{AKK}.

The authors of Ref.\cite{SS} studied intensively the D-branes in a
series of noncritical strings, the $(p,q)$
minimal conformal field theory coupled to two-dimensional quantum 
gravity (Liouville theory).\footnote{Here, $p$ and $q$ are
relatively prime integers and $p<q$.} They call this
the $(p,q)$ minimal string theory. They showed that there exist 
$\frac{(p-1)(q-1)}{2}$ independent ZZ branes labeled by
$(m,n)$, where $qm-pn>0$, and those ZZ branes correspond to the
singularities of an auxiliary Riemann surface that are formed
by the analytically continued boundary cosmological constant and the derivative
of the FZZT disk amplitude with respect to the boundary cosmological constant.
In particular, when $(p,q)=(2,2k-1)$, the minimal string theory is 
realized by the $k$-th multicritical region of 
the one matrix model \cite{MSS} and the ZZ branes are identified with 
the local extrema of the effective potential for a matrix eigenvalue
\cite{SS,HHIKKMT}. These extrema 
can be called the instantons. The annulus amplitudes between the 
D-branes in the minimal string theory were evaluated in \cite{KOPSS}
to examine the deformations by the D-branes.

The authors of Ref.\cite{HHIKKMT} explored in detail 
the nonperturbative effect stemming from the ZZ brane
in $c=0$ noncritical string theory (the $(2,3)$ minimal string theory)
from the viewpoints of the one matrix model as well as of the loop equations
(the string field theory).
One of the important results of Ref.\cite{HHIKKMT} is that the ratio of 
the 1-instanton sector in the partition function of the matrix model
to the 0-instanton 
sector is universal, namely, it does not depend on the detailed structure
in the potential of the matrix model. Actually, they confirmed this up to
next to leading order in the $1/N$ expansion by the explicit calculation.
This ratio is interpreted
as the chemical potential of the instanton.
The results of \cite{David} and \cite{AKK}
tell us that 
the leading order of this
ratio is equal to $e^{Z_{ZZ}}$,
where $Z_{ZZ}$ is the ZZ brane
disk amplitude and $Z_{ZZ}\sim 1/g_s$. The extension of this analysis
to the supersymmetric case ($\hat{c}=0$ type 0B string theory) was reported
very recently \cite{KKM}.

%In this paper, we turn on the interaction between the instantons in the one
%matrix model. 
In general,
the interaction between D-branes is given by the annulus amplitude between
the D-branes. If the identification of the instantons in the matrix model
with the ZZ branes in Liouville theory is valid, we should be able to
obtain the annulus
amplitudes between the ZZ branes by considering the interaction between
the instantons. This means that we go beyond the dilute gas approximation.
For this purpose, we need to consider the general 
$(2,2k-1)$ case. The reason is as follows.  
The interaction between the identical instantons diverges due to
the Vandermonde determinant for the matrix eigenvalues. Correspondingly,
the annulus amplitudes between the identical ZZ branes also diverges.
On the other hand, the annulus amplitudes between the different ZZ branes 
is finite.
The $(2,2k-1)$ theory possesses $k-1$ independent instantons in the one
matrix model and the $k-1$ independent ZZ branes in Liouville theory.
So, the $(2,3)$ case is not sufficient for our purpose.

First, we generalize the calculation of the ratio of the 1-instanton sector
to the 0-instanton sector
in \cite{HHIKKMT} to the $(2,2k-1)$ case. We show that it is also universal
for generic $k$ and the leading order is given by the ZZ brane disk amplitude.
Next, we consider the 2-instanton sector in the partition
function of the matrix model 
turning on the interaction between the instantons.
We show that the ratio of the 2-instanton sector to the 0-instanton sector
is universal and its connected part indeed reproduces
the annulus amplitudes between the ZZ branes in Liouville theory.
Our result serves as another nontrivial check on the correspondence 
between the instantons in the one matrix model and the ZZ branes 
in Liouville theory, 
and also suggests that
the expansion of the partition function in terms of the instanton numbers
are universal and gives systematically
the ZZ brane amplitudes in Liouville theory.

This paper is organized as follows. Sections 2 and 3 are devoted to the 
reviews of the ZZ branes in the $(2,2k-1)$ minimal string theory
and of the $k$-th multicritical
region of the one matrix model, respectively. In section 4, we see the 
behavior of the effective potential of a matrix eigenvalue and define
the expansion of the partition function in the instanton numbers.
Sections 5 and 6 are the main part of this paper.
We perform the above mentioned calculations of 
the ratio of the 1-instanton
sector to the 0-instanton sector in section 5 and of 
the ratio of the 2-instanton sector to the 0-instanton sector in section 6.
Section 7 is devoted to summary and discussion. 
In appendix A, we give the detailed 
calculation for the $(2,5)$ case. In appendix B, we 
gather some formulae used in the main text.

%%%%%%%%%%%%%%%%%%%%%%%%%%%%%%%%%%%%%%%%%%%%%%%%%%%%%%%%%%%%%%%%%%
\section{ZZ branes in Liouville theory}
%%%%%%%%%%%%%%%%%%%%%%%%%%%%%%%%%%%%%%%%%%%%%%%%%%%%%%%%%%%%%%%%%%
\setcounter{equation}{0}
In this section, we describe a part of the 
results in Refs.\cite{SS,KOPSS} which is
relevant for our purpose. Many of the features of the $(p,q)$ minimal 
string theory are provided by an auxiliary Riemann surface 
${\cal M}_{p,q}$, which is described by the algebraic equation
\beqa
F(\xi,\eta)=T_q(\xi)-T_p(\eta)=0,
\label{curve}
\eeqa
where $T_p(\cos\theta)=\cos p\theta$ is the Chebyshev polynomial of the
first kind. $\xi$ is the ratio of the boundary cosmological constant $\zeta$
to the square root of the bulk cosmological constant $\mu$ while $\eta$ is
proportional to the derivative of the FZZT disk amplitude $Z_{FZZT}$
with respect to $\zeta$:
\beqa
\xi=\frac{\zeta}{\sqrt{\mu}},\;\;\; 
\eta\sim \mu^{-\frac{q}{2p}}\pa_{\zeta}Z_{FZZT}.
\eeqa
It is convenient to introduce auxiliary parameters $\sigma$ and
$z=\cosh \frac{\pi\sigma}{\sqrt{pq}}$, in terms of which
\beqa
&&\xi=\cosh\pi\sqrt{\frac{p}{q}}\sigma=T_p(z),\n
&&\eta=\cosh\pi\sqrt{\frac{q}{p}}\sigma=T_{q}(z).
\eeqa
Note that $z$ covers the surface exactly once.
The ZZ branes correspond to the singularities of ${\cal M}_{p,q}$
given by $F=\pa_{\xi}F=\pa_{\eta}F=0$, which correspond to two
different values of $z$ denoted by $z^{\pm}$.

In what follows, we restrict ourselves to the case in which $(p,q)=(2,2k-1)$. 
In this case (\ref{curve}) is given by
\beqa
2\eta^2=T_{2k-1}(\xi)+1
=\frac{(T_k(\xi)+T_{k-1}(\xi))^2}{\xi+1}.
\eeqa
We define $\eta_k(\xi)$ by
\beqa
\sqrt{2}\eta_k(\xi)
=\frac{T_k(\xi)+T_{k-1}(\xi)}{\sqrt{\xi+1}}.
\label{eta}
\eeqa
Then, $F=\pa_{\xi}F=\pa_{\eta}F=0$ are equivalent to
\beqa
\eta_k=0,\;\; \pa_{\xi}\eta_k(\xi)^2=0.
\eeqa
These equations are solved as
\beqa
\xi_{n}=-\cos\frac{2\pi n}{2k-1}, \;\;\;\; n=1,2,\cdots,k-1,
\label{xin}
\eeqa
which corresponds to 
\beqa
z_n^{\pm}=-\sin\frac{\pm\pi n}{2k-1}.
\label{znpm}
\eeqa
$\xi_{n}$ characterizes the $(1,n)$ ZZ brane.
$\eta_k$ is expressed in terms of $\xi_n$'s as
\beqa
2\eta_k(\xi)^2
=2^{2k-2}(\xi-\xi_{1})^2(\xi-\xi_{2})^2\cdots
(\xi-\xi_{k-1})^2(\xi+1).
\eeqa
It is convenient to introduce an integral of $\eta_k$.
\beqa
v_k(\xi)\equiv\int^{\xi}d\xi'\sqrt{2}\eta_k(\xi')
=\frac{T_{k+1}(\xi)+T_k(\xi)}{(2k+1)\sqrt{\xi+1}}
-\frac{T_{k-1}(\xi)+T_{k-2}(\xi)}{(2k-3)\sqrt{\xi+1}}.
\label{vk}
\eeqa
This is proportional to the FZZT disk amplitude:
$v_k(\xi)\sim \mu^{-\frac{k}{2}-\frac{1}{4}}Z_{FZZT}$.
Later, we will use the following quantities which are proportional to
the ZZ brane disk amplitudes. 
\beqa
v_k(\xi_{n})=(-1)^{k+n}\sqrt{2}
\left(\frac{1}{2k+1}+\frac{1}{2k-3}\right)\sin\frac{2\pi n}{2k-1}.
\label{vkxin}
\eeqa
The annulus amplitudes between the ZZ branes were calculated in
\cite{KOPSS}. The result for that between the
$(1,n)$ and $(1,n')$ ZZ branes is
\beqa
Z_{n,n'}
=\log\frac{(z_n^+-z_{n'}^+)(z_n^--z_{n'}^-)}{(z_n^+-z_{n'}^-)(z_n^--z_{n'}^+)}.
\label{Znn'L}
\eeqa

%%%%%%%%%%%%%%%%%%%%%%%%%%%%%%%%%%%%%%%%%%%%%%%%%%%%%%%%%%%%%%%%%%%%
\section{The $k$-th multicritical region of the one matrix model}
%%%%%%%%%%%%%%%%%%%%%%%%%%%%%%%%%%%%%%%%%%%%%%%%%%%%%%%%%%%%%%%%%%%%
\setcounter{equation}{0}
We are concerned with the one matrix model with a generic potential.
\beqa
&&Z=\int d\phi \: e^{-N\tr V(\phi)}, \n
&&V(x)=\frac{1}{2}x^2-\sum_{m=3}^{\infty}\frac{g_m}{m}x^m,
\eeqa
where $\phi$ is an $N\times N$ Hermitian matrix. By diagonalizing $\phi$,
this integral is reduced to
\beqa
Z=\int \prod_{i=1}^{N}d\lambda_i \:
\Delta_N(\lambda_1,\cdots,\lambda_N)^2 e^{-N\sum_{i=1}^N V(\lambda_i)},
\label{partitionfuction}
\eeqa
where $\lambda_1,\cdots,\lambda_N$ are eigenvalues of $\phi$ and 
$\Delta_N(\lambda_1,\cdots,\lambda_N)$ is the Vandermonde determinant in terms
of $\lambda_1,\cdots,\lambda_N$. It is a standard technique to introduce
the orthogonal polynomials $P_n(x)$, which satisfy 
\beqa
\int dx \: e^{-NV(x)}P_m(x)P_n(x)=h_n\delta_{mn}, 
\eeqa
where $P_n(x)$ is a polynomial of degree $n$ and is
normalized so that the coefficient of $x^n$ equals one.  It is easy to
see that the following recursion relation holds.
\beqa
&&xP_n(x)=P_{n+1}(x)+s_nP_n(x)+r_nP_{n-1}(x), \\
&&r_n=\frac{h_n}{h_{n-1}}.
\eeqa
The partition
function $Z_N$ is expressed in terms of $r_n$:
\beqa
Z=N!h_0h_1\cdots h_{N-1}=N! h_0^N\prod_{n=1}^{N-1}r_n^{N-n}.
\label{Z}
\eeqa
Then, the relevant
part of the free energy $F=\log Z$ takes the form
\beqa
F=\sum_{n=1}^{N-1}(N-n)\log r_n.
\label{freeenergy}
\eeqa
%\beqa
%F=N^2\int_0^1 d\sigma \: (1-\sigma)\log r(\sigma)
%\eeqa
%where $\sigma=\frac{n}{N}$
The Schwinger-Dyson equations for the orthogonal polynomials,
\beqa
&&nh_{n-1}=\int dx\: e^{-NV(x)} \frac{dP_n(x)}{dx}P_{n-1}(x), \n
&&0=\int dx\: \frac{d}{dx}(e^{-NV(x)} P_n(x)P_{n}(x)),
\label{SDeqs}
\eeqa
give recursion relations for $r_n$ and $s_n$, from which one can 
determine $r_n$ and $s_n$ as functions of $g_m$'s.

We need the $k$-th multicritical region of the one matrix model to obtain
the $(2,2k-1)$ minimal string theory.
The $k$-th multicritical region is realized by
fine-tuning $k-1$ parameters among $g_m$'s and taking $N\rightarrow\infty$
limit. We introduce a continuous variable $\sigma=\frac{n}{N}$ in order to 
examine the critical behavior of the model. The critical point corresponds to
$\sigma=1$. As is explained in appendix C of 
Ref.\cite{HHIKKMT}, if the functions $r(\sigma)$ and $s(\sigma)$ are defined by
\beqa
r_n&=&r(\sigma), \n
s_n&=&s\left(\sigma+\frac{1}{2N}\right),
\label{rands}
\eeqa
$r(\sigma)$ and $s(\sigma)$ are ${\cal O}(N^0)$ quantities and the corrections
start with ${\cal O}(\frac{1}{N^2})$. Moreover, the ${\cal O}(N^0)$ parts
of $r(\sigma)$ and $s(\sigma)$ behave at the critical point like
\beqa
\frac{\frac{\pa r}{\pa \sigma}}{\frac{\pa s}{\pa \sigma}}=\sqrt{r_c},
\label{criticalbehaviorofrands}
\eeqa
where $r_c$ is the critical value of $r_n$.
Now we are ready to write down the scaling limit of the one matrix model
which gives rise to a perturbation around the $k$-th critical point and
corresponds to
the $(2,2k-1)$ minimal string theory \cite{MSS}:
\beqa
&&g_{m_i}=g_{m_i c}(1-\beta_{m_i}\mu\varepsilon^2),\;\;\;
i=1,\cdots,k-1, \n
&&r(\sigma)=r_c\left(1-\frac{1}{2}\alpha\varepsilon u(\tau)\right), \n
&&s(\sigma)=s_c-\frac{1}{2}\alpha\sqrt{r_c}\varepsilon u(\tau), \n
&&\sigma=1-\varepsilon^k\nu \tau, \n
&&\frac{1}{N}=\varepsilon^{k+\frac{1}{2}}\kappa g_s,
\label{doublescalinglimit}
\eeqa
where $\varepsilon$ is a cutoff so that
$\varepsilon\rightarrow 0$ corresponds to
the continuum limit. $g_{m_ic}$, $r_c$ and $s_c$ are critical values of
$g_{m_i}$, $r$ and $s$, respectively, which are dependent on
the detailed structure in the potential of the matrix model.
$\mu$ is the bulk cosmological constant which is
identified with
that in the Liouville theory. $\alpha$, $\beta_{m_i}$, $\nu$ and $\kappa$
are certain constants. 
In (\ref{doublescalinglimit}), we restrict ourselves to the leading order
of the $1/N$ expansion and have taken (\ref{criticalbehaviorofrands}) 
into account. 
$\alpha$, $\beta_{m_i}$ and $\nu$ in (\ref{doublescalinglimit}) 
are adjusted in such a way that $u(\sigma)$ 
obeys a string equation \cite{MSS}
\beqa
\sum_{j=0}^{\infty}t_j u^j=\tau,
\label{stringequation}
\eeqa
where
\beqa
&&t_{k-2p}=C_{k-2p} \mu^p, \;\;\;
p=0,1,\cdots, \left[\frac{k}{2}\right],\n
&&\qquad\qquad 
C_{k-2p}=\frac{(-1)^{k+1}\pi}{\sqrt{8}}
\frac{2^{k-2p}}{(k-2p)!p!\Gamma(p-k+\frac{3}{2})}, \n
&&\mbox{other}\; t_j=0.
\label{tj}
\eeqa
This represents the above mentioned perturbation around the $k$-th 
multicritical point.
When $\tau=0$, the string equation (\ref{stringequation}) allows a solution
\beqa
u(0)=\sqrt{\mu}.
\label{u(0)}
\eeqa

The universal part of the sphere contribution to the free energy 
(\ref{freeenergy}) is expressed by $u(\tau)$ as
\beqa
F^{(sphere)}=N^2\int_0^1d\sigma\: (1-\sigma)\log r(\sigma)
=\frac{1}{2}\kappa^{-2}\nu^2\alpha g_s^{-2}
\int^0d\tau\: \tau u(\tau).
\label{universal part of free energy}
\eeqa

In the following sections, we use the resolvent, which is defined in the 
large $N$ limit (the leading order of the $1/N$ expansion) by
\beqa
R(x)=\left\langle \frac{1}{N}\tr\left(\frac{1}{x-\phi}\right)\right\rangle.
\label{resolvent}
\eeqa
$R(x)$ is related to the eigenvalue density $\rho(x)$ as 
\beqa
\rho(x)=-\frac{1}{\pi}\mbox{Im} R(x+i0).
\eeqa
By solving the loop equation, the form of $R(x)$ is determined as
\beqa
&&R(x)=\frac{1}{2}V'(x)+W(x),\n
&&W(x)=\frac{1}{2}\sqrt{V'(x)^2+p(x)},
\label{RandW}
\eeqa
where $p(x)$ is a polynomial of degree $m_0-2$ when
\beqa
g_{m_0}\neq 0,\;\;\; g_m=0 \;\;\mbox{for} \;m>m_0.
\eeqa
$p(x)$ is determined by the structure of the cut, namely the location
where the eigenvalues are distributed, and the condition that 
$R(x)\sim\frac{1}{x}$ when $|x|\rightarrow\infty$.
We are interested in the one-cut solution, in which $\rho(x)$ is nonzero
only for the period $[b,a]$. Then $W(x)$ takes the form
\beqa
W(x)=\frac{1}{2}K(x)(x-x_1)(x-x_2)\cdots
(x-x_{k-1})\sqrt{(x-a)(x-b)},
\label{W}
\eeqa
where $K(x)$ is a polynomial of degree $m_0-k-1$. In the scaling limit
(\ref{doublescalinglimit}), $a$ and $x_n$ behave like
\beqa
a&=&x_*(1-\chi_a \sqrt{\mu}\varepsilon), \n
x_n&=&x_*(1+\chi_n \sqrt{\mu}\varepsilon),\;\;\; n=1,\cdots,k-1,
\eeqa
where $\chi_a$ and $\chi_n$ are given constants.
On the other hand, one can regard $b$ and $K(x)$ 
as some constants in the scaling
limit. If $x$ is scaled in the scaling limit (\ref{doublescalinglimit}) as
\beqa
x=x_*(1+\varepsilon\tilde{\zeta}),
\label{scalingofx}
\eeqa
$W(x)$ starts with a term proportional to $\varepsilon^{k-\frac{1}{2}}$ 
which is the universal part of the resolvent.
We will see in section 5
that $\chi_n=\xi_n\chi_a$ and $\chi_a=\alpha x_*^{-1}r_c^{\frac{1}{2}}$
for generic $V(x)$, $\chi_a=\frac{1}{4}\alpha$ for even $V(x)$.
Hence, if  $\tilde{\zeta}$ is tuned as $\tilde{\zeta}=\chi_a \zeta$,
the universal part of the resolvent becomes
proportional to the derivative of the FZZT disk amplitude with
respect to the boundary cosmological constant, namely $\eta(\xi)$. 
This is anticipated because the resolvent is interpreted as the expectation
value of a marked macroscopic loop in the matrix model and the macroscopic loop
is nothing but the geometrical meaning of the FZZT brane.
In appendix A, we illustrate the calculations
in this section with the case in which $k=3$ and the potential is even.
We see that
$\chi_n=\xi_n\chi_a$ and $\chi_a=\frac{1}{4}\alpha$ actually hold.

%Actually, in the scaling
%limit $NW(x)$ coincides with $\tilde{w}(\zeta)$ in Liouville theory that is 
%the Laplace transform of the marked disk amplitude $w(l)$:
%\beqa
%\tilde{w}(\zeta)=\int_0^{\infty}dl\: w(l)e^{-l\zeta}.
%\eeqa
%The marked disk amplitude was calculated in \cite{MSS} as
%\beqa
%w(l)=C_k g_s^{-1} l^{-1}\mu^{\frac{k}{2}-\frac{1}{4}}
%K_{k-\frac{1}{2}}(\sqrt{\mu}l),
%\eeqa
%where $l$ is the length of the boundary,
%$K_{k-\frac{1}{2}}$ is the modified Bessel function, and $C_k$ is a certain
%numerical constant.

%%%%%%%%%%%%%%%%%%%%%%%%%%%%%%%%%%%%%%%%%%%%%%%%%%%%%%%%%%%%%%%%%%%%%
\section{The effective potential for an eigenvalue and instantons}
%%%%%%%%%%%%%%%%%%%%%%%%%%%%%%%%%%%%%%%%%%%%%%%%%%%%%%%%%%%%%%%%%%%%%
\setcounter{equation}{0}
We consider the situation in which a single eigenvalue, say $\lambda_N$, is
separated from the others. The partition function (\ref{partitionfuction})
is expressed as
\beqa
Z = \int dx \int \prod_{i=1}^{N-1}d\lambda_i \:
\left(\prod_{i=1}^{N-1}(x-\lambda_i)^2\right) 
\Delta_{N-1}(\lambda_1,\cdots,\lambda_{N-1})^2\:
e^{-N\sum_{i=1}^{N-1} V(\lambda_i)}e^{-NV(x)},
\eeqa
where we set $\lambda_N=x$. By using an $(N-1)\times (N-1)$ Hermitian matrix
$\phi_{N-1}$, this is rewritten as
\beqa 
Z&=&Z_{N-1} \int dx \langle \det (x-\phi_{N-1})^2 \rangle_{N-1} e^{-NV(x)},
\eeqa
where
\beqa
Z_{N-1}&=&\int d\phi_{N-1}\: e^{-N\tr V(\phi_{N-1})},\n
\langle {\cal O} \rangle_{N-1}&=&\frac{1}{Z_{N-1}}\int d\phi_{N-1}
\: {\cal O}\:e^{-N\tr V(\phi_{N-1})} 
\eeqa
The effective potential for $x$ is defined by
\beqa
V_{eff}(x)=V(x)-\frac{1}{N}\log\langle \det (x-\phi_{N-1})^2 \rangle_{N-1}
\eeqa
in such a way that
\beqa
Z=Z_{N-1}\int dx\: e^{-NV_{eff}(x)}.
\eeqa
At leading order of the $1/N$ expansion, the following calculation 
is justified.
\beqa
\langle \det (x-\phi_{N-1})^2 \rangle_{N-1}
&=&\langle \det (x-\phi)^2 \rangle \n
&=&\exp\left[2\mbox{Re}\langle\tr\log(x-\phi)\rangle\right] \n
&=&\exp\left[ 2N\mbox{Re}\int^x dx' \:R(x') \right].
\eeqa
Therefore, using (\ref{RandW}), we find that
the leading order of $V_{eff}(x)$ is 
\beqa
V_{eff}^{(0)}(x)=-2\mbox{Re}\int^x dx'\:W(x').
\label{Veff0andW}
\eeqa
Or equivalently, ${V_{eff}^{(0)}}'(x)=-2\mbox{Re}W(x)$.
Then, we see from (\ref{W}) that $V_{eff}^{(0)}(x)$ is constant in the cut
and that $V_{eff}^{(0)}(x)$ takes local extrema at $x=x_1,\cdots, x_{k-1}$. 
We ignore extrema coming from $K(x)$ in (\ref{W}), since they
do not contribute in the scaling limit. In Fig. 1, we draw the shape of 
$V_{eff}^{(0)}(x)$ roughly in the $k=5$ case.

\begin{figure}[htbp]
\begin{center}
%\psfrag{cc}{$x_1$}
%\psfrag{dd}{$x_2$}
%\psfrag{ee}{$x_3$}
%\psfrag{ff}{$x_4$}
\includegraphics[height=5cm, keepaspectratio, clip]{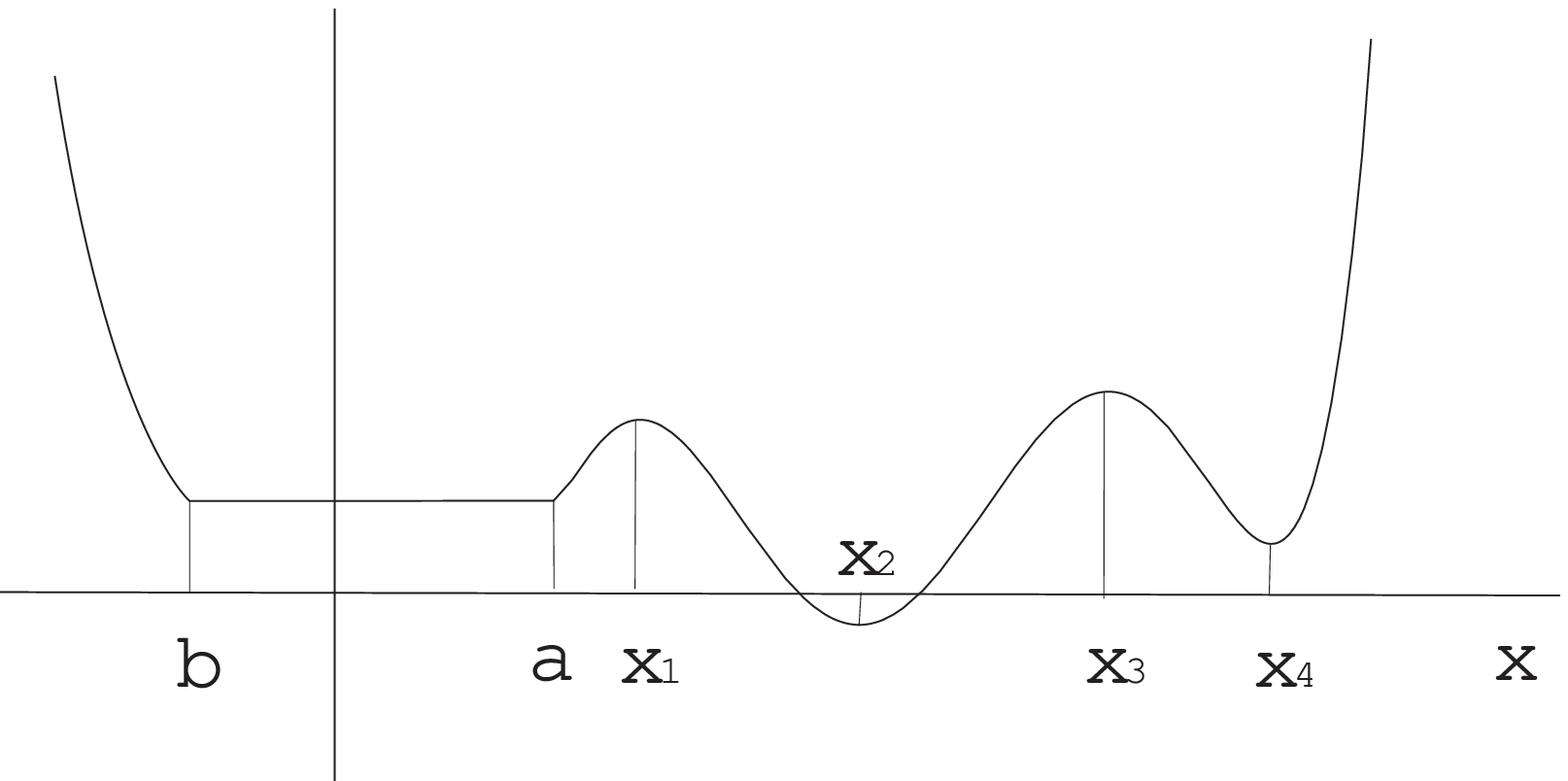}
\end{center}
\caption{Effective potential for an eigenvalue in the $k=5$ case}
\label{boundary}
\end{figure}

These extrema can be considered as the instantons in the one matrix model
and will be identified with the ZZ branes in Liouville theory.
We label by $\{q_1,\cdots,q_{k-1}\}$
the configuration in which $q_n\;(n=1,\cdots,k-1)$ eigenvalues among $N$ 
are located
around $x=x_n$  and the other $N-q$ eigenvalues
are located in the cut, where $q=q_1+\cdots +q_{k-1}$.
Namely, $q_n$ is the instanton
number of the $n$-th instanton. We denote by $\int_{x_n} dx$ the 
perturbative expansion around the `classical solution' $x=x_n$, which yields
a $1/N$ expansion. The leading and subleading contributions to
this expansion are nothing but the saddle point integral over $x$ around
$x=x_n$. 
We expand the partition function in terms of the instanton numbers as follows.
\beqa
Z&=&\sum_{q_1,\cdots,q_{k-1}=0}^{\infty} Z^{\{q_1,\cdots,q_{k-1}\}}\n
&=&Z^{(0-inst.)}\sum_{q_1,\cdots,q_{k-1}=0}^{\infty}
{\cal A}^{\{q_1,\cdots,q_{k-1}\}},
\label{expansion}
\eeqa
where
\beqa
Z^{\{q_1,\cdots,q_{k-1}\}}
&=&\frac{N!}{q_1!\cdots q_{k-1}!(N-q)!}
\left(\prod_{n=1}^{k-1}\int_{x_n}\prod_{i_n=1}^{q_n}dx_{i_n}^{(n)}\right)
\int_{b\leq\lambda_i\leq a}\prod_i^{N-q}d\lambda_{i=1} \n
&&\qquad\;\;\;\times
\Delta_{q}(x_1^{(1)},\cdots,x_{q_1}^{(1)},x_1^{(2)},\cdots,x_{q_2}^{(2)},
\cdots,x_1^{(k-1)},\cdots,x_{q_{k-1}}^{(k-1)})^2 \n
&&\qquad\;\;\;\times\left(\prod_{n=1}^{k-1}\prod_{i_n=1}^{q_n}\prod_{i=1}^{N-q}
(x_{i_n}^{(n)}-\lambda_i)^2\right) \:
\Delta_{N-q}(\lambda_1,\cdots,\lambda_{N-q})^2 \n
&&\qquad\;\;\;\times e^{-N\sum_{i=1}^{N-q}V(\lambda_i)} \:
e^{-N\sum_{n=1}^{k-1}\sum_{i_n=1}^{q_n}V(x_{i_n}^{(n)})}, \n
Z^{(0-inst.)}&=&Z^{\{0,\cdots,0\}}
=\int_{b\leq\lambda_i\leq a}\prod_{i=1}^{N}d\lambda_i\:
\Delta_N(\lambda_1,\cdots,\lambda_N)^2\: e^{-N\sum_{i=1}^N V(\lambda_i)},  \n
{\cal A}^{\{q_1,\cdots,q_{k-1}\}}&=&
\frac{Z^{\{q_1,\cdots,q_{k-1}\}}}{Z^{(0-inst.)}}, \n
{\cal A}^{\{0,\cdots,0\}}&=&1.
\eeqa
For example,
\beqa
&&Z^{\{0,\cdots,0,q_n=1,0,\cdots,0,q_{n'}=1,0,\cdots,0\}} \n
&&=N(N-1)\int_{x_n}dx\int_{x_{n'}}dy \int_{b\leq\lambda_i\leq a}
\prod_{i=1}^{N-2}d\lambda_i \n
&&\qquad\;\;\;\;\;\;\;\;\;\times (x-y)^2 
\left(\prod_{i=1}^{N-2}(x-\lambda_i)^2(y-\lambda_i)^2\right)
\Delta_{N-2}(\lambda_1,\cdots,\lambda_{N-2})^2 \n
&&\qquad\;\;\;\;\;\;\;\;\;
\times e^{-N\sum_{i=1}^{N-2}V(\lambda_i)}\:e^{-NV(x)-NV(y)}.
\eeqa
We make the following abbreviations for the quantities which we are
concerned with in subsequent sections.
\beqa
&&{\cal A}^{\{0,\cdots,0,q_n=1,0,\cdots,0\}}
=\frac{Z^{\{0,\cdots,0,q_n=1,0,\cdots,0\}}}{Z^{(0-inst.)}}
={\cal A}^{(n)},\n
&&{\cal A}^{\{0,\cdots,0,q_n=2,0,\cdots,0\}}
=\frac{Z^{\{0,\cdots,0,q_n=2,0,\cdots,0\}}}{Z^{(0-inst.)}}
={\cal A}^{(n,n)},\n
&&{\cal A}^{\{0,\cdots,0,q_n=1,0,\cdots,0,q_{n'}=1,0,\cdots,0\}}
=\frac{Z^{\{0,\cdots,0,q_n=1,0,\cdots,0,q_{n'}=1,0,\cdots,0\}}}{Z^{(0-inst.)}}
={\cal A}^{(n,n')}.
\eeqa
${\cal A}^{(n)}$ is the ratio of the 1-instanton sector to the 0-instanton 
sector while ${\cal A}^{(n,n)}$ and ${\cal A}^{(n,n')}$ are the ratios
of the 2-instanton sectors to the 0-instanton sector. We make 
${\cal A}^{(n,n')}$ represent both the cases, $n=n'$ and $n\neq n'$. 
The $1/N$ expansion on which our calculations in the following sections are 
based is not the expansion in term of $1/N^2$ but the one
in terms of $1/N$ due to the instanton effects. It will turn out that
$\log{\cal A}^{(n)}$ and $\log{\cal A}^{(n,n')}$ start with 
${\cal O}(N)$. We will evaluate ${\cal O}(N)$, ${\cal O}(\log N)$ and
${\cal O}(N^0)$ terms, which become ${\cal O}(1/g_s)$, 
${\cal O}(\log g_s)$ and ${\cal O}(g_s^0)$ terms in the continuum limit,
respectively.

%%%%%%%%%%%%%%%%%%%%%%%%%%%%%%%%%%%%%%%%%%%%%%%%%%%%%%%%%%%%%%%%%%%%%%
\section{1-instanton sectors and the ZZ brane disk amplitudes}
%%%%%%%%%%%%%%%%%%%%%%%%%%%%%%%%%%%%%%%%%%%%%%%%%%%%%%%%%%%%%%%%%%%%%%
\setcounter{equation}{0}
In this section, we calculate ${\cal A}^{(n)}$. $\log{\cal A}^{(n)}$ will 
turn out to start with ${\cal O}(N)$. 
We will evaluate ${\cal O}(N)$, ${\cal O}(\log N)$
and ${\cal O}(N^0)$ terms in $\log{\cal A}^{(n)}$. We will show that these
terms are universal and the leading order term (the ${\cal O}(N)$ term) 
agrees with
the $(1,n)$ ZZ brane disk amplitude. 
We first write down the definition of ${\cal A}^{(n)}$.
\beqa
{\cal A}^{(n)}&=&\frac{Z^{\{0,\cdots,0,q_n=1,0,\cdots,0\}}}{Z^{(0-inst.)}} 
\n
&=&\frac{1}{Z^{(0-inst.)}}N
\int_{x_n}dx\int_{b\leq\lambda_i\leq a}\prod_{i=1}^{N-1}
d\lambda_i\:\prod_{i=1}^{N-1}(x-\lambda_i)^2
\Delta_{N-1}(\lambda_1,\cdots,\lambda_{N-1})^2\n
&&\qquad\qquad\qquad\qquad\qquad\qquad \times
e^{-N\sum_{i=1}^{N-1}V(\lambda_i)}\:e^{-NV(x)} \n
&=&N\frac{Z_{N-1}^{(0-inst.)}}{Z^{(0-inst.)}}
\int_{x_n}dx\: \langle\det(x-\phi_{N-1})^2\rangle_{N-1}^{(0-inst.)}\:
e^{-NV(x)},
\label{An}
\eeqa
where 
\beqa
Z_{N-1}^{(0-inst.)}&=&\int_{b\leq\lambda_i\leq a}\prod_{i=1}^{N-1}d\lambda_i\:
\Delta_{N-1}(\lambda_1,\cdots,\lambda_{N-1})^2\: 
e^{-N\sum_{i=1}^{N-1} V(\lambda_i)}, \n
\langle\det(x-\phi_{N-1})^2\rangle_{N-1}^{(0-inst.)}
&=&\frac{1}{Z_{N-1}^{(0-inst.)}}
\int_{b\leq\lambda_i\leq a}\prod_{i=1}^{N-1}d\lambda_i\:
\prod_{i=1}^{N-1}(x-\lambda_i)^2
\Delta_{N-1}(\lambda_1,\cdots,\lambda_{N-1})^2\n
&&\qquad\qquad\qquad\qquad
 \times e^{-N\sum_{i=1}^{N-1}V(\lambda_i)}, \n
&=&\frac{\int_{0-inst.}d\phi_{N-1}\:\det(x-\phi_{N-1})^2\:
e^{-N\tr V(\phi_{N-1})}}
{\int_{0-inst.}d\phi_{N-1}\:e^{-N\tr V(\phi_{N-1})}}.
\label{ZN-1anddetN-1}
\eeqa
Here $\phi_{N-1}$ is an $(N-1)\times (N-1)$ Hermitian matrix.

Keeping (\ref{Z}) in mind, we can calculate 
the factor $Z_{N-1}^{(0-inst.)}/Z^{(0-inst.)}$ in the last line of
(\ref{An})
as follows.
\beqa
\frac{Z_{N-1}^{(0-inst.)}}{Z^{(0-inst.)}}
%=\frac{(N-1)!h_0h_1\cdots h_{N-2}}{N!h_0h_1\cdots h_{N-1}}
=\frac{1}{Nh_{N-1}}\left(1+{\cal O}\left(\frac{1}{N}\right)\right)
=\frac{r_N}{Nh_N}\left(1+{\cal O}\left(\frac{1}{N}\right)\right).
\label{factor}
\eeqa
The ${\cal O}(1/N)$ correction comes from the instanton contributions.
The calculation of $h_N$ in appendix E of 
Ref.\cite{HHIKKMT} holds for our case. The result is 
\beqa
h_N=2\pi\sqrt{r(1)}e^{-NV_{eff}^{(0)}(a)}
\left(1+{\cal O}\left(\frac{1}{N}\right)\right).
\label{hN}
\eeqa
We need $\langle\det(x-\phi_{N-1})^2\rangle_{N-1}^{(0-inst.)}$
outside the cut, $x>a$.
By setting $x=\lambda_N$ formally, we find
\beqa
&&\langle\det(x-\phi_{N-1})^2\rangle_{N-1}^{(0-inst.)} \n
&&=\frac{1}{(N-1)!h_0h_1\cdots h_{N-2}}
\int_{b\leq\lambda_i\leq a}\prod_{i=1}^{N-1}d\lambda_i\:
\Delta_N(\lambda_1,\cdots,\lambda_N)^2\: e^{-N\sum_{i=1}^{N-1} V(\lambda_i)} \n
&&=P_{N-1}(\lambda_N)^2+\frac{h_{N-1}}{h_{N-2}}P_{N-2}(\lambda_N)^2+\cdots
+\frac{h_{N-1}}{h_0}P_0(\lambda_N)^2 \n
&&=P_{N-1}(x)^2+r_{N-1}P_{N-2}(x)^2+\cdots+r_{N-1}r_{N-2}\cdots r_{1}P_0(x)^2.
\label{detN-1andP}
\eeqa
(\ref{detN-1andP}) would hold exactly
if there was no limitation to the 0-instanton sector. Actually, there is
a relative ${\cal O}(1/N)$ error coming from the instanton contributions 
in each term in the last line
of (\ref{detN-1andP}). Here we ignored these errors
because they turn out to
only lead to a relative ${\cal O}(1/N)$ correction in the final result.
(\ref{detN-1andP}) is further
calculated following Ref.\cite{HHIKKMT}.
\beqa
&&\langle\det(x-\phi_{N-1})^2\rangle_{N-1}^{(0-inst.)} \n
&=&\left(\frac{k^{(0)}(x,1)}{q(x,1)}\right)^2
\exp\left[2N\int_0^{1-\frac{1}{N}}d\sigma\: \log k^{(0)}(x,\sigma)\right]
\left(1+{\cal O}\left(\frac{1}{N}\right)\right) \n
&=&\left(\frac{k^{(0)}(x,1)}{q(x,1)}\right)^2
\frac{1}{(k^{(0)}(x,1))^2}
\exp\left[2N\int_0^1d\sigma\: \log k^{(0)}(x,\sigma)\right]
\left(1+{\cal O}\left(\frac{1}{N}\right)\right), 
\label{detN-1}
\eeqa
where
\beqa
&&k^{(0)}(x,\sigma)=\frac{1}{2}\left(x-s(\sigma)
+\sqrt{(x-s(\sigma))^2-4r(\sigma)}\right), \n
&&q(x,\sigma)=\sqrt{(x-s(\sigma))^2-4r(\sigma)}.
\label{k0andq}
\eeqa
These are ${\cal O}(N^0)$ quantities.
(\ref{detN-1}) implies that outside the cut
\beqa
V_{eff}^{(0)}(x)=V(x)-2\int_0^1 d\sigma \: \log k^{(0)}(x,\sigma).
\label{Veff0}
\eeqa
This would coincide with (\ref{Veff0andW}) outside the cut, so that
the structure of the cut in (\ref{Veff0}) should agree with that in $W(x)$.
This observation leads to a relation
\beqa
(x-s(1))^2-4r(1)=(x-a)(x-b),
\label{rsab}
\eeqa
from which we obtain 
\beqa
x_*=s_c+2\sqrt{r_c}.
\label{x*scrc}
\eeqa
$\mbox{}$From (\ref{An}), (\ref{factor}) and (\ref{detN-1}), we obtain
\beqa
{\cal A}^{(n)}=\frac{1}{h_N}\int_{x_n}dx\:\frac{r_N}{(k^{(0)}(x,1))^2}
\left(\frac{k^{(0)}(x,1)}{q(x,1)}\right)^2 \: e^{-NV_{eff}^{(0)}(x)}
\left(1+{\cal O}\left(\frac{1}{N}\right)\right),
\label{An2}
\eeqa
where $h_N$ and $V_{eff}^{(0)}(x)$ are given in (\ref{hN}) and (\ref{Veff0}),
respectively. From this expression, we see that $\log{\cal A}^{(n)}$
starts with ${\cal O}(N)$ and the saddle point integral
in (\ref{An2}) actually
gives ${\cal O}(N)$, ${\cal O}(\log N)$ and ${\cal O}(N^0)$
terms in $\log{\cal A}^{(n)}$.
%\footnote{(\ref{An2}) with $k=2$ and $n=1$ differs from 
%the expression for $\mu$ in
%Ref.\cite{HHIKKMT} by the factor $r_N/(k^{(0)}(x,1))^2$ in the integrand.
%However, this factor turns out not to contribute to the final result.}

First, we assume that $V(x)$ is generic without accidental symmetry.
We must treat separately the case
in which $V(x)$ is even. Substituting (\ref{doublescalinglimit}) and
(\ref{scalingofx}) into
the derivative of (\ref{Veff0}) and using (\ref{x*scrc}) leads to
\beqa
{V_{eff}^{(0)}}'(x)=V'(x)-\int_0^{\varepsilon^{-k}\nu^{-1}}d\tau
\: \varepsilon^{k-\frac{1}{2}}\nu
\left(\frac{1}{\sqrt{r_c\alpha u+r_c^{\frac{1}{2}}x_*\tilde{\zeta}}}
+{\cal O}(\varepsilon^{\frac{1}{2}})\right).
\label{Veff02}
\eeqa
As mentioned in section 3, $W(x)=-\frac{1}{2}{V_{eff}^{(0)}}'(x)$ 
starts with a term proportional to 
$\varepsilon^{k-\frac{1}{2}}$ in the scaling limit,
so that we are allowed to simplify (\ref{Veff02}) as 
\beqa
{V_{eff}^{(0)}}'(x)=\nu\alpha^{-\frac{1}{2}}r_c^{-\frac{1}{2}}
\varepsilon^{k-\frac{1}{2}}
\int^0 d\tau\:
\frac{1}{\sqrt{u(\tau)+\alpha^{-1}r_c^{-\frac{1}{2}}x_*\tilde{\zeta}}}
+{\cal O}({\epsilon^k}),
\label{Veff02.5}
\eeqa
where we 
keep only the contribution from the one end $\tau=0$ of the integral
region. The other terms with integer power
in $\varepsilon$ lower than $\varepsilon^{k-\frac{1}{2}}$ would be canceled
in (\ref{Veff02}). $\pa_{\tilde{\zeta}}V_{eff}^{(0)}(x)$ starts with the 
${\cal O}(\varepsilon^{k+\frac{1}{2}})$ term 
and $N\sim \varepsilon^{-k-\frac{1}{2}}$,
so that $N\pa_{\tilde{\zeta}}V_{eff}^{(0)}(x)$ is finite and we can ignore
the ${\cal O}(\varepsilon^k)$ correction in (\ref{Veff02.5}).
We also perform the change of the integration variable $u(\tau)=\sqrt{\mu}w$.
Then, using (\ref{stringequation}) and (\ref{u(0)}), we obtain
\beqa
\frac{\pa V_{eff}^{(0)}(x)}{\pa\tilde{\zeta}}
=\varepsilon^{k+\frac{1}{2}}\nu\alpha^{-\frac{1}{2}}r_c^{-\frac{1}{2}}
x_*\sum_{j=1}^{\infty}jt_j\mu^{\frac{j}{2}-\frac{1}{4}}
\int^1dw\: w^{j-1}
\frac{1}
{\sqrt{w+\alpha^{-1}r_c^{-\frac{1}{2}}x_*\mu^{-\frac{1}{2}}\tilde{\zeta}}}.
\label{Veff03}
\eeqa
%\beqa
%\left(\frac{\pa u}{\pa s}\right)^{-1}=\sum_{j=1}^{\infty}jt_ju^{j-1}
%\eeqa
Substituting (\ref{tj}) into (\ref{Veff03}) and using the formula (\ref{Pn})
yields
\beqa
\frac{\pa V_{eff}^{(0)}(x)}{\pa\tilde{\zeta}}
=\varepsilon^{k+\frac{1}{2}}\sqrt{\frac{\pi}{2}}\nu\alpha^{\frac{1}{2}}\Omega
\mu^{\frac{k}{2}-\frac{1}{4}}\int_{-\frac{\Omega\tilde{\zeta}}{\sqrt{\mu}}}^1 
dw\:
\left(w+\frac{\Omega\tilde{\zeta}}{\sqrt{\mu}}\right)^{-\frac{1}{2}}P_{k-1}(w),
\eeqa
where $\Omega=\alpha^{-1}r_c^{-\frac{1}{2}}x_*$ and $P_{k-1}$ is the
Legendre polynomial of degree $k-1$. We have specified consistently
the lower end of the integral region in such a way that it does
not contribute to the value of the integral.
Furthermore, using 
(\ref{Pn-}), (\ref{formula}) and (\ref{eta}), we obtain
\beqa
\frac{\pa V_{eff}^{(0)}(x)}{\pa\tilde{\zeta}}
=\varepsilon^{k+\frac{1}{2}}\nu\alpha^{\frac{1}{2}}\Omega
\frac{(-1)^{k+1}\sqrt{2\pi}}{2k-1}\mu^{\frac{k}{2}-\frac{1}{4}}
\sqrt{2}\eta_k(\Omega\mu^{-\frac{1}{2}}\tilde{\zeta}).
\label{Veff04}
\eeqa
The above calculation that reduces (\ref{Veff03}) to (\ref{Veff04}) is
essentially same as that in appendix B in \cite{SS}.
The lefthand side of (\ref{Veff04}) is  proportional to
the leading term in the scaling limit of $W(x)$, which is the universal part
of the resolvent, so that we obtain 
\beqa
\chi_a=\Omega^{-1}=\alpha x_*^{-1}r_c^{\frac{1}{2}},
\;\;\; \chi_n=\xi_n\chi_a.
\label{chiaandchin}
\eeqa
That is, $x=x_n$ corresponds to 
$\tilde{\zeta}=\Omega^{-1}\mu^{\frac{1}{2}}\xi_n$ and $x=a$ corresponds to
$\tilde{\zeta}=-\Omega^{-1}\mu^{\frac{1}{2}}$.
By integrating (\ref{Veff04}) over $\tilde{\zeta}$, we finally obtain
\beqa
NV_{eff}^{(0)}(x)=(\kappa^{-1}\nu\alpha^{\frac{1}{2}})
\frac{(-1)^{k+1}\sqrt{2\pi}}{2k-1}g_s^{-1}\mu^{\frac{k}{2}+\frac{1}{4}}
v_k(\Omega\mu^{-\frac{1}{2}}\tilde{\zeta}).
\label{NVeff0}
\eeqa

In order to determine the overall factor $\kappa^{-1}\nu\alpha^{\frac{1}{2}}$ 
in (\ref{NVeff0}),
we need a physical input. 
We adopt the sphere amplitude as the physical input. First, 
we calculate
the sphere contribution to the free energy of the matrix model
(\ref{universal part of free energy}). By performing the change of the variable
$u(\tau)=\sqrt{\mu}w$ and using (\ref{stringequation}), (\ref{tj}),
(\ref{u(0)}), (\ref{Pn}) and (\ref{formula2}),
we obtain
\beqa
F^{(sphere)}&=&\frac{1}{2}(\kappa^{-1}\nu\alpha^{\frac{1}{2}})^2 g_s^{-2}
\mu^{k+\frac{1}{2}}
\sum_{p=0}^{[\frac{k-1}{2}]}\sum_{q=0}^{[\frac{k}{2}]}(k-2p)C_{k-2p}C_{k-2q}
\int^{1}dw\: w^{2k-2p-2q} \n
&=&\frac{1}{2}(\kappa^{-1}\nu\alpha^{\frac{1}{2}})^2\Xi_k g_s^{-2} 
\mu^{k+\frac{1}{2}},
\label{F}
\eeqa
where
\beqa
\Xi_k
%=\sum_{p=0}^{[\frac{k-1}{2}]}\sum_{q=0}^{[\frac{k}{2}]}
%\frac{k-2p}{2k-2p-2q+1}C_{k-2p}C_{k-2q}
=-\frac{\pi}{2}\frac{1}{(2k+1)(2k-1)(2k-3)}.
\label{Xi}
\eeqa
If the sphere amplitude is given by 
\beqa
F^{(sphere)}=d_kg_s^{-2}\mu^{k+\frac{1}{2}},
\label{Ffixed}
\eeqa
where $d_k$ is a certain universal constant,
$\kappa^{-1}\nu\alpha^{\frac{1}{2}}$ is fixed as
\beqa
\kappa^{-1}\nu\alpha^{\frac{1}{2}}=\sqrt{\frac{2d_k}{\Xi_k}}.
\eeqa
Then, $NV_{eff}^{(0)}(x)$ is determined as
\beqa
NV_{eff}^{(0)}(x)=D_k
g_s^{-1}\mu^{\frac{k}{2}+\frac{1}{4}}
v_k(\Omega\mu^{-\frac{1}{2}}\tilde{\zeta}),
\label{NVeff0universal}
\eeqa
where
\beqa
D_k=\frac{2\sqrt{\pi}(-1)^{k+1}}{2k-1}\sqrt{\frac{d_k}{\Xi_k}}
=(-1)^{k+1}2\sqrt{2}\sqrt{\frac{(2k+1)(2k-3)}{2k-1}}\sqrt{-d_k}.
\label{Dk}
\eeqa
%It is nontrivial that once a physical quantity, the sphere
%amplitude, is given the normalization of $NV_{eff}^{(0)}(x)$ is fixed
%universally.

We are ready to calculate $A^{(n)}$ given in (\ref{An2}).
\beqa
{\cal A}^{(n)}&=&\frac{1}{h_N}\frac{r_N}{(k^{(0)}(x_n,1))^2}
\left(\frac{k^{(0)}(x_n,1)}{q(x_n,1)}\right)^2 
x_*\varepsilon
e^{-D_kg_s^{-1}\mu^{\frac{k}{2}+\frac{1}{4}}v_k(\xi_n)} \n
&&\qquad\times\int d\tilde{\zeta}\: 
e^{-\frac{D_kg_s^{-1}}{2}
\mu^{\frac{k}{2}-\frac{3}{4}}\Omega^2 v_k''(\xi_n)
(\tilde{\zeta}-\Omega^{-1}\mu^{\frac{1}{2}}\xi_n)^2} \n
&=&\frac{1}{h_N}\frac{r_N}{(k^{(0)}(x_n,1))^2}
\left(\frac{k^{(0)}(x_n,1)}{q(x_n,1)}\right)^2 
x_*\Omega^{-1}\varepsilon\sqrt{\frac{2\pi}{D_kv_k''(\xi_n)}}
g_s^{\frac{1}{2}}\mu^{-\frac{k}{4}+\frac{3}{8}}
e^{-D_kg_s^{-1}\mu^{\frac{k}{2}+\frac{1}{4}}v_k(\xi_n)}, \n
\label{An3}
\eeqa
where we have used (\ref{scalingofx}). We see from (\ref{hN}), (\ref{x*scrc}) 
and (\ref{chiaandchin}) that
in the scaling limit
\beqa
&&h_N=2\pi \sqrt{r_c} e^{-D_kg_s^{-1}\mu^{\frac{k}{2}+\frac{1}{4}}v_k(-1)}
=2\pi \sqrt{r_c} \n
&&\frac{r_N}{(k^{(0)}(x_n,1))^2}=1, \n
&&\left(\frac{k^{(0)}(x_n,1)}{q(x_n,1)}\right)^2 
=\frac{1}{4\varepsilon\alpha\sqrt{\mu}}\frac{1}{\xi_n+1}.
\label{somequantities}
\eeqa
Thus, noting $\Omega=\alpha^{-1}r_c^{-\frac{1}{2}}x_*$, we finally obtain
\beqa
{\cal A}^{(n)}=\frac{1}{8}\sqrt{\frac{2}{\pi D_kv_k''(\xi_n)}}
\frac{1}{\xi_n+1}g_s^{\frac{1}{2}}\mu^{-\frac{k}{4}-\frac{1}{8}}
e^{-D_kg_s^{-1}\mu^{\frac{k}{2}+\frac{1}{4}}v_k(\xi_n)},
\label{Anfinal}
\eeqa
where $\xi_n$, $v_k(\xi)$, $v_k(\xi_n)$ and $D_k$ are given in
(\ref{xin}), (\ref{vk}), (\ref{vkxin}) and (\ref{Dk}), respectively.
It follows from this expression that ${\cal A}^{(n)}$ 
is indeed universal i.e. independent of the detailed structure in the potential of the matrix model.

Next, let us consider the case in which $V(x)$ is even. 
$s_n$ vanishes identically in this case. We can see that 
all the equations (\ref{Veff02})$\sim$(\ref{NVeff0}) also hold 
for this case if we set $x_*=2\sqrt{r_c}$ and replace $\alpha$ 
with $\frac{\alpha}{2}$. 
Namely,
\beqa
NV_{eff}^{(0)}(x)=(\kappa^{-1}\nu\alpha^{\frac{1}{2}})
\frac{(-1)^{k+1}\sqrt{\pi}}{2k-1}g_s^{-1}\mu^{\frac{k}{2}+\frac{1}{4}}
v_k(\Omega\mu^{-\frac{1}{2}}\tilde{\zeta})
\label{NVeff0even}
\eeqa
with $\Omega=4\alpha^{-1}$ and $\chi_a=\frac{1}{4}\alpha$.
$F^{(sphere)}$ given in (\ref{universal part of 
free energy}) does not include $s_n$ so that (\ref{F}) is invariant.
However, instead of (\ref{Ffixed}) we must fix $F^{(sphere)}$ as
\beqa
F^{(sphere)}=2d_kg_s^{-2}\mu^{k+\frac{1}{2}},
\eeqa
since due to the $Z_2$ symmetry there are two critical points each of which
contributes equally to the free energy. 
Hence, $\kappa^{-1}\nu\alpha^{\frac{1}{2}}$ in (\ref{NVeff0even}) 
is determined
as
\beqa
\kappa^{-1}\nu\alpha^{\frac{1}{2}}=2\sqrt{\frac{d_k}{\Xi_k}}.
\eeqa
This indeed reduces (\ref{NVeff0even}) to (\ref{NVeff0universal}).
(\ref{An3}) and (\ref{somequantities}) also hold for this case if we 
set $x_*=2\sqrt{r_c}$ and replace $\alpha$ with $\frac{\alpha}{2}$. 
Noting $\Omega=4\alpha^{-1}$, we can easily see that $A^{(n)}$ is indeed
given by (\ref{Anfinal}). 

Thus, the proof of the universality of $A^{(n)}$
is completed. Note that $A^{(n)}$ is pure imaginary for odd $n$ 
and real for even $n$. This reflects the fact that $\xi=\xi_n$ with
$n$ odd corresponds to a local maximum and $\xi=\xi_n$ with $n$ even
corresponds to a local minimum.

As a check on our calculation, let us calculate $A^{(1)}$ in the $k=2$ case,
which would coincide with $\mu$ in \cite{HHIKKMT}. 
The normalization of $F^{(sphere)}$ in \cite{HHIKKMT} corresponds to 
$d_2=-\frac{4}{15}$.
Using (\ref{xin}), (\ref{vk}), (\ref{vkxin})
and (\ref{Dk}), we can calculate the quantities that appears in (\ref{Anfinal})
with $k=2$ and $n=1$ as
\beqa
D_2=-\frac{4\sqrt{2}}{3},\;\;\;
\xi_1=\frac{1}{2},\;\;\; v_2(\xi_1)=-\frac{3\sqrt{6}}{5},\;\;\;
v_2''(\xi_1)=\sqrt{6}.
\eeqa
By substituting these quantities into (\ref{Anfinal}), we obtain
\beqa
{\cal A}^{(1)}=\frac{i}{8\cdot 3^{\frac{3}{4}}\sqrt{\pi}}g_s^{\frac{1}{2}}
\mu^{-\frac{5}{8}}e^{-\frac{8\sqrt{3}}{5}g_s^{-1}\mu^{\frac{5}{4}}}.
\eeqa
This indeed coincides with $\mu$ in \cite{HHIKKMT}.

Finally, let us see that the leading order of $\log{\cal A}^{(n)}$
indeed agrees with
the $(1,n)$ ZZ brane disk amplitude, which we denote by $Z_n$.
The leading order of $\log{\cal A}^{(n)}$ is given by
\beqa
-D_k g_s^{-1} \mu^{\frac{k}{2}+\frac{1}{4}}v_k(\xi_n).
\label{leadingorderofAn}
\eeqa
We can evaluate $Z_n$ in Liouville theory
by using eqs.(B.4) and (B.6) in \cite{KOPSS} as
\beqa
Z_n=(-1)^n \frac{4\cdot 2^{\frac{5}{4}}}{\pi^{\frac{3}{2}}}
\frac{(2k-1)^{\frac{1}{4}}}{2k+1}
\frac{\Gamma(\frac{2k-3}{2k-1})}{\Gamma(\frac{2k-1}{2})}
\left(\sin\frac{2\pi}{2k-1}\right)^{\frac{1}{2}}
\sin\frac{2\pi n}{2k-1} g_s^{-1}\mu^{\frac{k}{2}+\frac{1}{4}}.
\label{Zn}
\eeqa
We can also calculate the sphere amplitude in Liouville theory 
by integrating twice
the two-point function of the cosmological constant operators which is
given in (2.26) in \cite{AKK}. The result corresponds to\footnote{The relation
between our cosmological constant $\mu$ and the cosmological constant
$\mu_L$ in \cite{AKK} is 
$\mu=\mu_L\pi \frac{\Gamma(\frac{2k-3}{2k-1})}{\Gamma(\frac{2}{2k-1})}$.
Note also that for odd $k$ the overall sign of $d_k$ obtained from
(2.26) in \cite{AKK}
is different from that in (\ref{dkL}). It seems possible to attribute
this difference to the ambiguity
of the sign of the norm in nonunitary models. We adjust the overall
sign of $d_k$ in (\ref{dkL}) to minus in such a way that it is 
consistent with the result
in the matrix model.}
\beqa
d_k=-\frac{1}{\pi^3}\sqrt{\frac{2}{2k-1}}\frac{2k-3}{2k+1}
\sin\frac{2\pi}{2k-1}
\left(\frac{\Gamma(\frac{2k-3}{2k-1})}{\Gamma(\frac{2k-1}{2})}\right)^2.
\label{dkL}
\eeqa
It is easy to verify
that plugging (\ref{dkL}) into (\ref{leadingorderofAn}) actually
yields (\ref{Zn}).

The fact that the leading order of $\log{\cal A}^{(n)}$ is the $(1,n)$ ZZ
brane disk amplitude is already pointed out in \cite{SS}.
What is new in this section is that we showed that both the leading order term
(${\cal O}(1/g_s)$)
and, in particular, the next to leading order terms (${\cal O}(\log g_s)$ and  
${\cal O}(g_s^0)$) in $\log{\cal A}^{(n)}$ are universal.
We also showed that the normalization of the ZZ
brane disk amplitude is also reproduced precisely by matching 
the sphere amplitude in the matrix model with that in Liouville theory.

%%%%%%%%%%%%%%%%%%%%%%%%%%%%%%%%%%%%%%%%%%%%%%%%%%%%%%%%%%%%%%%%
\section{2-instanton sectors and the annulus
amplitudes between the ZZ branes}
%%%%%%%%%%%%%%%%%%%%%%%%%%%%%%%%%%%%%%%%%%%%%%%%%%%%%%%%%%%%%%%%
\setcounter{equation}{0}
In this section, we calculate ${\cal A}^{(n,n')}$. The estimation of 
the order in the $1/N$ expansion proceeds in the same way as that in the
calculation of ${\cal A}^{(n)}$. We will not dwell on it in this section.
$\log{\cal A}^{(n,n')}$
also starts with ${\cal O}(N)$. We will evaluate 
${\cal O}(N)$, ${\cal O}(\log N)$
and ${\cal O}(N^0)$ terms in $\log{\cal A}^{(n,n')}$.
We will see that these terms are universal. We will also show that
in the $n\neq n'$ case
the leading order term
in $\log{\cal A}^{(n,n')}_c$, which is ${\cal O}(N^0)$, reproduces
the annulus amplitude between the $(1,n)$ and $(1,n')$ ZZ branes, where
${\cal A}^{(n,n')}_c={\cal A}^{(n,n')}/{\cal A}^{(n)}{\cal A}^{(n')}$.
First, we consider the $n\neq n'$ case.
In this case ${\cal A}^{(n,n')}$ is given by
\beqa
{\cal A}^{(n,n')}&=&
\frac{Z^{\{0,\cdots,0,q_n=1,0,\cdots,0,q_{n'}=1,0,\cdots,0\}}}
{Z^{(0-inst.)}}\n
&=&\frac{1}{Z^{(0-inst.)}}
N(N-1)\int_{x_n}dx\int_{x_{n'}}dy \int_{b\leq\lambda_i\leq a}
\prod_{i=1}^{N-2}d\lambda_i \n
&&\qquad\qquad\;\;\;\times (x-y)^2 
\left(\prod_{i=1}^{N-2}(x-\lambda_i)^2(y-\lambda_i)^2\right)
\Delta_{N-2}(\lambda_1,\cdots,\lambda_{N-2})^2 \n
&&\qquad\qquad\;\;\;\times e^{-N\sum_{i=1}^{N-2}V(\lambda_i)}\:
e^{-NV(x)-NV(y)} \n
&=&N(N-1)\frac{Z_{N-2}^{(0-inst.)}}{Z^{(0-inst.)}}
\int_{x_n}dx\int_{x_{n'}}dy\:
\langle\det(x-\phi_{N-2})^2\det(y-\phi_{N-2})^2\rangle_{N-2}^{(0-inst.)} \n
&&\qquad\qquad\qquad\qquad\qquad \times e^{-NV(x)-NV(y)+\log(x-y)^2},
\label{Ann'}
\eeqa
where 
\beqa
&&Z_{N-2}^{(0-inst.)}=\int_{b\leq\lambda_i\leq a}\prod_{i=1}^{N-2}d\lambda_i\:
\Delta_{N-2}(\lambda_1,\cdots,\lambda_{N-2})^2\: 
e^{-N\sum_{i=1}^{N-2} V(\lambda_i)}, \n
&&
\langle\det(x-\phi_{N-2})^2\det(y-\phi_{N-2})^2\rangle_{N-2}^{(0-inst.)}
\n
&&=\frac{1}{Z_{N-2}^{(0-inst.)}}
\int_{b\leq\lambda_i\leq a}\prod_{i=1}^{N-2}d\lambda_i\:
\left(\prod_{i=1}^{N-2}(x-\lambda_i)^2(y-\lambda_i)^2\right)
\Delta_{N-2}(\lambda_1,\cdots,\lambda_{N-2})^2\n
&&\qquad\qquad\qquad\qquad\qquad\qquad
 \times e^{-N\sum_{i=1}^{N-2}V(\lambda_i)}, \n
&&=\frac{\int_{0-inst.}d\phi_{N-2}\:
\det(x-\phi_{N-2})^2\det(y-\phi_{N-2})^2\:
e^{-N\tr V(\phi_{N-2})}}
{\int_{0-inst.}d\phi_{N-2}\:e^{-N\tr V(\phi_{N-2})}}.
\label{ZN-2anddetN-2}
\eeqa
Here $\phi_{N-2}$ is an $(N-2)\times (N-2)$ Hermitian matrix.
The factor $Z_{N-2}^{(0-inst.)}/Z^{(0-inst.)}$ in the last line
of (\ref{Ann'}) is calculated as
\beqa
\frac{Z_{N-2}^{(0-inst.)}}{Z^{(0-inst.)}}
=\frac{1}{N(N-1)h_{N-1}h_{N-2}}
=\frac{r_N^2r_{N-1}}{N(N-1)h_N^2}.
\label{factor2}
\eeqa
$\langle\det(x-\phi_{N-2})^2\det(y-\phi_{N-2})^2\rangle_{N-2}^{(0-inst.)}$ 
is evaluated as follows.
\beqa
&&
\langle\det(x-\phi_{N-2})^2\det(y-\phi_{N-2})^2\rangle_{N-2}^{(0-inst.)}
\n
&&=\langle e^{\tr\log(x-\phi_{N-2})^2+\tr\log(y-\phi_{N-2})^2}
\rangle_{N-2}^{(0-inst.)}\n
&&=\exp\left[
\langle\tr\log(x-\phi_{N-2})^2\rangle_{N-2}^{(0-inst.)}
+\langle\tr\log(y-\phi_{N-2})^2\rangle_{N-2}^{(0-inst.)}\right.\n
&&\qquad\qquad
+\frac{1}{2}\langle(\tr\log(x-\phi_{N-2})^2)^2\rangle_{N-2,c}^{(0-inst.)}
+\frac{1}{2}\langle(\tr\log(y-\phi_{N-2})^2)^2\rangle_{N-2,c}^{(0-inst.)}
\n
&&\qquad\qquad
\left.+\langle\tr\log(x-\phi_{N-2})^2\tr\log(y-\phi_{N-2})^2
\rangle_{N-2,c}^{(0-inst.)}+\cdots\right] \n
&&=\langle\det(x-\phi_{N-2})^2\rangle_{N-2}^{(0-inst.)}
\langle\det(y-\phi_{N-2})^2\rangle_{N-2}^{(0-inst.)} \n
&&\;\;\;\;\;\times\exp\left[\langle
\tr\log(x-\phi_{N-2})^2\tr\log(y-\phi_{N-2})^2
\rangle_{N-2,c}^{(0-inst.)}+\cdots\right],
\label{detN-2detN-2}
\eeqa
where the subscript `$c$' stands for the connected part.
We can calculate $\langle\det(x-\phi_{N-2})^2\rangle_{N-2}^{(0-inst.)}$ 
in a way similar to (\ref{detN-1andP}) and (\ref{detN-1}) as
\beqa
&&\langle\det(x-\phi_{N-1})^2\rangle_{N-2}^{(0-inst.)} \n
%&=&\left(\frac{k^{(0)}(x,1)}{q(x,1)}\right)^2
%\exp\left[2N\int_0^{1-\frac{2}{N}}d\sigma\: \log k^{(0)}(x,\sigma)\right] \n
&&=\left(\frac{k^{(0)}(x,1)}{q(x,1)}\right)^2
\frac{1}{(k^{(0)}(x,1))^4}
\exp\left[2N\int_0^1d\sigma\: \log k^{(0)}(x,\sigma)\right].
\label{detN-2}
\eeqa
A similar expression holds for
$\langle\det(y-\phi_{N-2})^2\rangle_{N-2}^{(0-inst.)}$.
In the leading order of the $1/N$ expansion, we can set
\beqa
\langle\tr\log(x-\phi_{N-2})^2\tr\log(y-\phi_{N-2})^2
\rangle_{N-2,c}^{(0-inst.)}
=4\langle\tr\log(x-\phi)\tr\log(y-\phi)\rangle_c.
\label{cylinder}
\eeqa
The righthand side is calculated in appendix B. From (\ref{Ann'}), 
(\ref{factor2}), (\ref{detN-2detN-2}), (\ref{detN-2}) and (\ref{cylinder}),
we obtain
\beqa
&&{\cal A}^{(n,n')}
=\frac{r_N^2r_{N-1}}{h_N^2}\int_{x_n}dx\int_{x_{n'}}dy\:
\left(\frac{k^{(0)}(x,1)}{q(x,1)}\right)^2\frac{1}{(k^{(0)}(x,1))^4}
\left(\frac{k^{(0)}(y,1)}{q(y,1)}\right)^2\frac{1}{(k^{(0)}(y,1))^4}
\n
&&\qquad
\times\exp\left[-NV_{eff}^{(0)}(x)-NV_{eff}^{(0)}(y)+\log(x-y)^2
+4\langle\tr\log(x-\phi)\tr\log(y-\phi)\rangle_c\right]
\eeqa
By noting that in the scaling limit
\beqa
r_N^2r_{N-1}\left(\frac{1}{k^{(0)}(x_n,1)}\right)^4
\left(\frac{1}{k^{(0)}(x_{n'},1)}\right)^4
=\frac{1}{r_c}
\eeqa
and recalling the calculation of ${\cal A}^{(n)}$, we find
\beqa
{\cal A}^{(n,n')}&=&{\cal A}^{(n)}{\cal A}^{(n')}{\cal A}^{(n,n')}_c, \n
{\cal A}^{(n,n')}_c&=&
\exp\left[4\langle\tr\log(x_n-\phi)\tr\log(x_{n'}-\phi)\rangle_c
+\log(x_n-x_{n'})^2-\log r_c\right]. 
\eeqa
Here  ${\cal A}^{(n,n')}_c$ is interpreted
as the `connected part' of ${\cal A}^{(n,n')}$ since
${\cal A}^{(n)}{\cal A}^{(n')}$ is the product of the 1-instanton
contributions. As seen in the previous
section, the leading order
of $\log{\cal A}^{(n)}$ is the $(1,n)$ ZZ brane disk amplitude, so
that the leading order of $\log{\cal A}^{(n,n')}_c$ is expected to be
the annulus amplitude between the $(1,n)$ and $(1,n')$ ZZ branes.
In the following, we will show that this is indeed the case. 
Using (\ref{trlntrln}) leads to
\beqa
&&\log {\cal A}^{(n,n')}_c \n
&&=4\langle\tr\log(x_n-\phi)\tr\log(x_{n'}-\phi)\rangle_c
+\log(x_n-x_{n'})^2-\log r_c \n
&&=\log\left(2x_nx_{n'}-(a+b)(x_n+x_{n'})+a^2+b^2
-2\sqrt{(x_n-a)(x_n-b)(x_{n'}-a)(x_{n'}-b)}\right)^2 \n
&&\;\;\;\;-\log\left(\sqrt{(x_n-a)(x_n-b)}+\sqrt{(x_{n'}-a)(x_{n'}-b)}
\right)^4 \n
&&\;\;\;\;+\log\left(\sqrt{x_n-a}+\sqrt{x_n-b}\right)^4
+\log\left(\sqrt{x_{n'}-a}+\sqrt{x_{n'}-b}\right)^4
-\log(a-b)^4-\log 16 \n
&&\;\;\;\;+\log(x_n-x_{n'})^2-\log r_c
\label{lnAnn'c1}
\eeqa
Recalling that in the scaling limit,
\beqa
a&=&x_*(1-\chi_a\sqrt{\mu}\varepsilon), \n
x_n&=&x_*(1+\chi_a\xi_n\sqrt{\mu}\varepsilon),
\label{scalingofaandxn}
\eeqa
we calculate the quantities that appears in (\ref{lnAnn'c1}).
\beqa
&&2x_nx_{n'}-(a+b)(x_n+x_{n'})+a^2+b^2
-2\sqrt{(x_n-a)(x_n-b)(x_{n'}-a)(x_{n'}-b)} \n
&&=(x_*-b)^2(1+{\cal O}(\varepsilon)), \n
&&\sqrt{(x_n-a)(x_n-b)}+\sqrt{(x_{n'}-a)(x_{n'}-b)}
=\sqrt{x_*-b}\left(\sqrt{x_n-a}+\sqrt{x_{n'}-a}\right)
(1+{\cal O}(\varepsilon^{\frac{1}{2}})), \n
&&\sqrt{x_n-a}+\sqrt{x_n-b}
=\sqrt{x_*-b}\:(1+{\cal O}(\varepsilon^{\frac{1}{2}})),\n
&&\sqrt{x_{n'}-a}+\sqrt{x_{n'}-b}
=\sqrt{x_*-b}(1+{\cal O}\:(\varepsilon^{\frac{1}{2}})), \n
&&a-b=(x_*-b)(1+{\cal O}(\varepsilon)).
\eeqa
Substituting these into (\ref{lnAnn'c}) yields
\beqa
\log {\cal A}^{(n,n')}_c
=\log\left[\frac{(x_n-x_{n'})^2}{(\sqrt{x_n-a}+\sqrt{x_{n'}-a})^4}
\frac{(x_*-b)^2}{16r_c}\right]+{\cal O}(\varepsilon^{\frac{1}{2}}).
\eeqa
Furthermore, by using (\ref{scalingofaandxn}) and a relation
\beqa
r_c=\frac{(x_*-b)^2}{16}(1+{\cal O}(\varepsilon))
\eeqa
which follows from (\ref{rsab}), we finally obtain
\beqa
\log {\cal A}^{(n,n')}_c
=\log\frac{(\xi_n-\xi_{n'})^2}{(\sqrt{\xi_n+1}+\sqrt{\xi_{n'}+1})^4}.
\label{lnAnn'c}
\eeqa
This is universal. The model-dependent quantities such as $r_c$ are indeed
canceled. It follows that ${\cal A}^{(n,n')}$ is also universal because
${\cal A}^{(n)}$ and ${\cal A}^{(n')}$ are universal.

Let us see that $\log{\cal A}^{(n,n')}$ with $n\neq n'$
is the annulus amplitude between
the ZZ branes in Liouville theory.
First, we rewrite $z_n^{\pm}$ by $\xi_n$.
\beqa
z_n^{\pm}=-\sin\frac{\pm\pi n}{2k-1}=\mp\sqrt{\frac{\xi_n+1}{2}}.
\eeqa
Using this, we express the annulus amplitude between the $(1,n)$ and $(1,n')$
ZZ branes (\ref{Znn'L}) in terms of $\xi_n$:
\beqa
Z_{n,n'}=\log\frac{(\xi_n-\xi_{n'})^2}{(\sqrt{\xi_n+1}+\sqrt{\xi_{n'}+1})^4}.
\eeqa
This indeed agrees with $\log {\cal A}^{(n,n')}_c$.

Next, we consider the case in which $n=n'$. The same calculation as
the $n\neq n'$ case leads to
\beqa
{\cal A}^{(n,n)}&=&\frac{1}{2}({\cal A}^{(n)})^2
\exp\left[4\langle\tr\log(x_n-\phi)\tr\log(x_n-\phi)\rangle_c
+\log(x_n-x_n)^2-\log r_c\right] \n
&=&\frac{1}{2}({\cal A}^{(n)})^2
\frac{(\xi_n-\xi_n)^2}{(\sqrt{\xi_n+1}+\sqrt{\xi_n+1})^4}.
\eeqa
That is, ${\cal A}^{(n,n)}$ vanishes and ${\log\cal A}^{(n,n)}$ diverges.
This is consistent with the result in Liouville theory where $Z_{n,n}$ also
diverges.\footnote{It is already pointed out in \cite{KOPSS} that 
the double zero of $e^{Z_{nn}}$ stems from the 
Vandermonde determinant of the matrix model.}

%%%%%%%%%%%%%%%%%%%%%%%%%%%%%%%%%%%%%%%%%%%%%%%%%%%%%%%%%%%%%%%%%%
\section{Summary and discussion}
%%%%%%%%%%%%%%%%%%%%%%%%%%%%%%%%%%%%%%%%%%%%%%%%%%%%%%%%%%%%%%%%%%
\setcounter{equation}{0}
In this paper, we analyzed the $k$-th multicritical region of 
the one matrix model that corresponds to the $(2,2k-1)$ minimal string
theory. We divided the partition function of the matrix model
in terms of the instanton numbers.  We evaluated the ratio of the 1-instanton
sector to the 0-instanton sector, ${\cal A}^{(n)}$, and the ratio of
the 2-instanton sector to the 0-instanton sector, ${\cal A}^{(n,n')}$.
We found that
$\log{\cal A}^{(n)}$ and $\log{\cal A}^{(n,n')}$ start with ${\cal O}(N)$
terms.
We calculated the ${\cal O}(N)$, ${\cal O}(\log N)$ and ${\cal O}(N^0)$ terms
in $\log{\cal A}^{(n)}$ and $\log{\cal A}^{(n,n')}$, which correspond to
${\cal O}(1/g_s)$, ${\cal O}(\log g_s)$ and ${\cal O}(g_s^0)$ terms 
in the continuum limit, respectively.
Our results are as follows.
(i)${\cal O}(N)$, ${\cal O}(\log N)$ and ${\cal O}(N^0)$ terms
in $\log{\cal A}^{(n)}$ and $\log{\cal A}^{(n,n')}$ 
are universal i.e. independent
of the detailed structure in the potential of the matrix model. 
(ii)The ${\cal O}(N)$ term in $\log{\cal A}^{(n)}$ is equal to the $(1,n)$
ZZ brane disk amplitude, which is proportional to $1/g_s$. 
(iii)When ${\cal A}^{(n,n')}$ with $n\neq n'$ is expressed as 
${\cal A}^{(n,n')}={\cal A}^{(n)}{\cal A}^{(n')}{\cal A}^{(n,n')}_c$,
$\log{\cal A}^{(n,n')}_c$ starts with ${\cal O}(N^0)$ term. This term
reproduces the annulus amplitude between the $(1,n)$ and $(1,n')$ ZZ branes
in Liouville theory. 
(iv)The ${\cal O}(N)$ term in $\log{\cal A}^{(n,n)}$ are given by that
in $\log{\cal A}^{(n)}$ while one of the ${\cal O}(N^0)$ terms 
in $\log{\cal A}^{(n,n)}$
diverges. This makes $\log{\cal A}^{(n,n)}$ vanish and is consistent with
the result in Liouville theory.

The above results allow us to assign Figs. 2(a) and 4 to 
the leading order of $\log{\cal A}^{(n)}$ and $\log{\cal A}^{(n,n')}_c$,
respectively. We express ${\cal A}^{(n,n)}$ as
${\cal A}^{(n,n)}=\frac{1}{2}({\cal A}^{(n)})^2{\cal A}^{(n,n)}_c$.
Then, it is natural to assign Fig. 2(b) to the next to leading
order (${\cal O}(\log N)$ and ${\cal O}(N^0)$) of $\log{\cal A}^{(n)}$ 
and Fig. 3 to the leading order of
${\cal A}^{(n,n)}_c$, which is divergent.  Namely, the next to leading
order of $\log{\cal A}^{(n)}$ can be interpreted as
the annulus stretched within a single $(1,n)$ ZZ brane while
the leading order of $\log{\cal A}^{(n,n)}$
can be interpreted as the annulus stretched from 
one $(1,n)$ ZZ brane to the other $(1,n)$ ZZ brane.
In Liouville theory, these two diagrams are not distinguished because
the ZZ branes have no intrinsic parameter such as the position except
the label $n$, so that the annulus amplitude between 
the two identical $(1,n)$ ZZ 
branes in Liouville theory is consistently divergent. As is stressed in
\cite{HHIKKMT} in the $k=2$ case, a nontrivial thing
is that the next to leading order of $\log{\cal A}^{(n)}$ is a finite and
universal quantity, which cannot be evaluated at least so far in Liouville
theory. ${\cal A}^{(n)}$ has a physical interpretation as
the chemical potential of the $n$-th instanton, so that
it is suggested that the matrix model possesses the information on
the nonperturbative effect that Liouville theory cannot predict.
Note that as shown in \cite{HHIKKMT} this quantity cannot be
calculated through the loop equation (the string field theory), either.

Our results imply that ${\cal A}^{\{q_1,\cdots,q_{k-1}\}}$ vanishes 
if $q_n \geq 2$ at least for a certain $n$ so that the expansion of the
partition function is terminated with $2^{k-1}$ terms as 
\beqa
Z=Z^{(0-inst.)}\sum_{q_1,\cdots,q_{k-1}=0,1}{\cal A}^{\{q_1,\cdots,q_{k-1}\}}.
\label{truncatedZ}
\eeqa
The maximum of the total instanton number is $q=k-1$.
Actually, ${\cal A}^{\{1,1,\cdots,1\}}$ is nonvanishing.
Our results suggest that each ${\cal A}^{\{q_1,\cdots,q_{k-1}\}}$ in
(\ref{truncatedZ}) is universal and its `connected part' systematically
reproduces the amplitudes among the ZZ branes in Liouville theory.
It is interesting to calculate multi-point amplitudes among the ZZ branes
in Liouville theory if possible and compare them with
${\cal A}^{\{q_1,\cdots,q_{k-1}\}}$ in (\ref{truncatedZ}).
The reason why ${\cal A}^{\{q_1,\cdots,q_{k-1}\}}$ with $q_n \geq 2$ 
at least for a certain $n$ vanishes 
is that one cannot place more than one eigenvalue at the same point
because of the repulsive force coming from the Vandermonde determinant.
However, taking into account $\log(x_n-x_{n'})^2$ stemming from
the Vandermonde determinant from the beginning, 
one can obtain
a `classical solution' that realizes the situation in which
those eigenvalues are separated a little from each other
around a local extreme of the potential.
Perhaps one can construct 
an expansion of the partition function
based on this classical solution, which differs from the expansion
in this paper.
This expansion would lead to the
singularity destroying deformation argued in \cite{KOPSS}, which
changes the singularity to a cut, and may yield ${\cal O}((e^{Z_n})^m)$
$(m\geq 2)$ corrections to (\ref{truncatedZ}). Note that in the $k=2$ case
one can reproduce from (\ref{truncatedZ}) with these possible corrections
the prediction of the string equation
that the deviation from the perturbative solution for
the free energy $F=\log Z$ behaves as $\sim e^{Z_1}$ at leading order
of $e^{Z_1}$,
because at this order the logarithm of $Z$ in (\ref{truncatedZ}) with the
possible corrections takes
the form $F=F^{(0-inst.)}+{\cal A}^{(1)}$.

%On the other hand, in the dilute gas approximation
%the interaction between the
%instantons is turned off. In other words, one ignores
%the factor $(x_n-x_{n'})^2$ stemming
%from the Vandermonde determinant and so on. Hence, in this approximation,
%the expansion of the partition function obviously reduces to
%\beqa
%Z&=&Z^{(0-inst.)}\sum_{q=0}^{\infty}\sum_{q_1+\cdots+q_{k-1}=q}
%\frac{1}{q_1!\cdots q_{k-1}!}
%({\cal A}^{(1)})^{q_1}\cdots ({\cal A}^{(k-1)})^{q_{k-1}} \n
%&=&Z^{(0-inst.)}e^{{\cal A}^{(1)}+\cdots+{\cal A}^{(k-1)}}.
%\eeqa
%Or equivalently, 
%\beqa
%F=F^{(0-inst.)}+{\cal A}^{(1)}+\cdots+{\cal A}^{(k-1)}.
%\label{fullF}
%\eeqa
%This means that the chemical potential of the $n$-th instanton is 
%${\cal A}^{(n)}$. 

Finally, we make a comment.
As is pointed in \cite{SS,KOPSS}, 
the sign of $NV_{eff}^{(0)}(\xi_n)$ is $(-1)^{n+1}$ so that
${\cal A}^{(n)}$ with $n$ even
behaves as $\sim e^{+\frac{1}{g_s}}$, which is catastrophic.
The energy of $n$-th instanton with $n$ even which corresponds to
the local minimum is below the Fermi level. 
Therefore, the perturbative vacuum is unstable
due to the eigenvalues tunneling to these local minima.
This is due to the nonunitary nature of the model. Note that the fact that
in the $(2,3)$ case $n$ takes only 1 is consistent with the unitarity
of the $(2,3)$ model. 
Thus, the expansion of the partition function of the one matrix model
in the instanton numbers should be understood as a formal one in this sense.

It is important to generalize our analysis to the two matrix
model, which can represent the unitary noncritical string theories.

\begin{figure}[htbp]
\begin{center}
\includegraphics[height=2.5cm, keepaspectratio, clip]{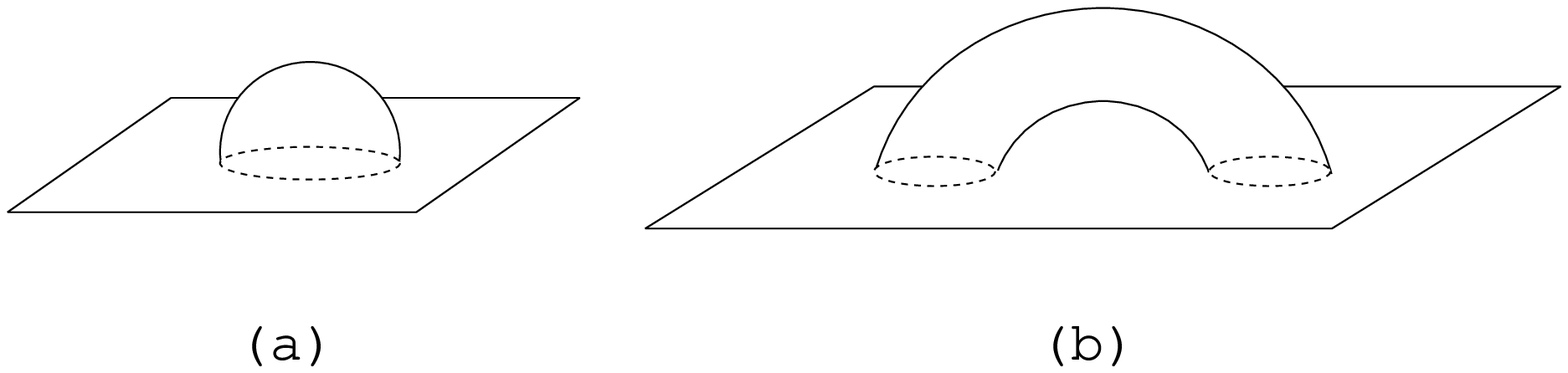}
\end{center}
\caption{$\log{\cal A}^{(n)}$: (a)the leading order (b)the next to 
leading order}
\label{lnAn}
\end{figure}

\begin{figure}[htbp]
\begin{center}
\includegraphics[height=2cm, keepaspectratio, clip]{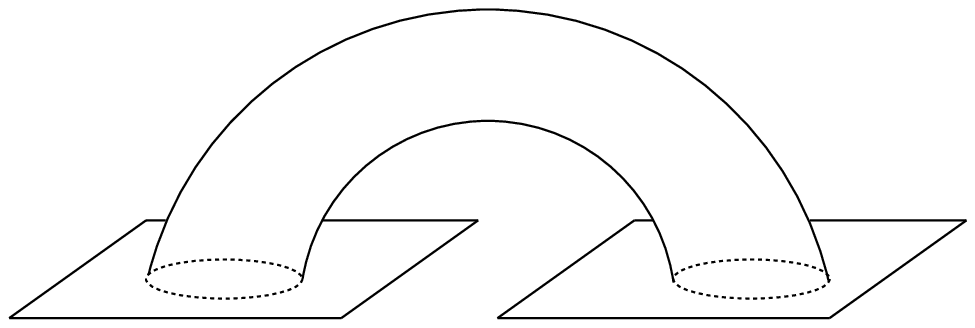}
\end{center}
\caption{The leading order of $\log{\cal A}^{(n,n)}_c$}
\label{lnAnn}
\end{figure}

\begin{figure}[htbp]
\begin{center}
\includegraphics[height=2cm, keepaspectratio, clip]{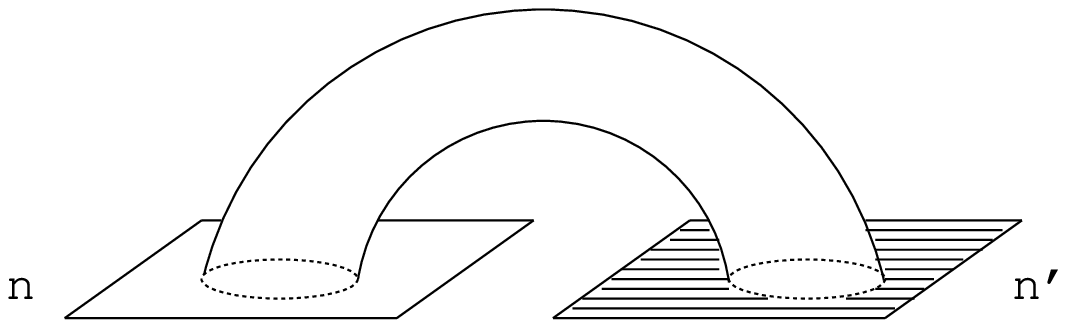}
\end{center}
\caption{The leading order of $\log{\cal A}^{(n,n')}_c$}
\label{lnAnn'}
\end{figure}

\section*{Acknowledgements}
We would like to thank H. Kawai and T. Nakatsu for discussions.
The work of A.T. is supported in part by Grant-in-Aid for Scientific
Research (No.16740144) from the Ministry of Education, Science and Culture.

%%%%%%%%%%%%%%%%%%%%%%%%%%%%%%%%%%%%%%%%%%%%%%%%%%%%%%%%%%%%%%%%%%
\section*{Appendix A: The $(2,5)$ model}
%%%%%%%%%%%%%%%%%%%%%%%%%%%%%%%%%%%%%%%%%%%%%%%%%%%%%%%%%%%%%%%%%%
\setcounter{equation}{0}
\renewcommand{\theequation}{A.\arabic{equation}}
In this appendix, as an example, we analyze in detail the one matrix
model in the 3rd critical region ($k=3$),
which corresponds to the $(2,5)$ minimal string theory.
We adopt an even potential of the one matrix model, 
\beqa
V(x)=\frac{1}{2}x^2-\frac{1}{4}g_4 x^4 -\frac{1}{6}g_6 x^6.
\eeqa
The first equation in (\ref{SDeqs}) gives in the large-$N$ limit
\beqa
\sigma=r(\sigma)-3g_4 r(\sigma)^2-10g_6 r(\sigma)^3,
\label{k=3recursionrelation}
\eeqa
while the second one is trivial. Correspondingly, $s(\sigma)$ vanishes
identically. The scaling limit (\ref{doublescalinglimit}) is given 
in this case by
\beqa
&&g_4=\frac{1}{9}(1-\beta\mu\varepsilon^2), \n
&&g_6=-\frac{1}{270}(1-3\beta\mu\varepsilon^2), \n
&&r(\sigma)=3\left(1-\frac{1}{2}\alpha\varepsilon u(\tau)\right), \n
&&\sigma=1-\varepsilon^3 \nu \tau, \n
&&\frac{1}{N}=\varepsilon^{\frac{7}{2}}\kappa g_s.
\label{k=3scalinglimit}
\eeqa
Plugging (\ref{k=3scalinglimit}) into (\ref{k=3recursionrelation}) gives
\beqa
\tau
=-\frac{3\beta\alpha}{2\nu}\mu u+\frac{\alpha^3}{8\nu}u^3.
\eeqa
Comparing this equation with the string equation (\ref{stringequation}),
we find
\beqa
\frac{\alpha^3}{8\nu}=\sqrt{\frac{\pi}{8}}, \;\;\;
\frac{3\beta\alpha}{2\nu}=\sqrt{\frac{\pi}{8}},
\eeqa
which are equivalent to
\beqa
\alpha=2\sqrt{3}\beta^{\frac{1}{2}},\;\;\;
\nu=\sqrt{\frac{8}{\pi}}3\sqrt{3}\beta^{\frac{3}{2}}.
\label{alphaandnu}
\eeqa
%\beqa
%F=-\frac{4}{105}g_s^{-2}\kappa^{-2}(9\beta\mu)^{\frac{7}{2}}
%\eeqa
%\beqa
%R(x)=\frac{1}{2}V'(x)+W(x)
%\eeqa
$W(x)$ takes the form
\beqa
W(x)=\frac{1}{2}g_6(x^2-x_1^2)(x^2-x_2^2)\sqrt{x^2-a^2}.
\eeqa
The condition that $R(x)\sim\frac{1}{x}$ when $x\rightarrow\infty$ is 
equivalent to
\beqa
&&g_4+g_6\left(\frac{1}{2}a^2+x_1^2+x_2^2\right)=0, \n
&&1-g_6\left(\frac{1}{8}a^4-\frac{1}{2}a^2(x_1^2+x_2^2)-x_1^2x_2^2\right)=0, \n
&&-g_6\left(\frac{1}{32}a^6-\frac{1}{16}a^4(x_1^2+x_2^2)
+\frac{1}{4}a^2x_1^2x_2^2\right)=1.
\eeqa
In the scaling limit (\ref{k=3scalinglimit}), $a$, $x_1$ and $x_2$ are 
determined by these equations as
\beqa
a&=&x_*\left(1-\frac{\sqrt{3}}{2}\sqrt{\beta\mu}\varepsilon\right), \n
x_1&=&x_*\left(1+\frac{\sqrt{3}}{2}\frac{1-\sqrt{5}}{4}
\sqrt{\beta\mu}\varepsilon\right), \n
x_2&=&x_*\left(1+\frac{\sqrt{3}}{2}\frac{1+\sqrt{5}}{4}
\sqrt{\beta\mu}\varepsilon\right),
\eeqa
where $x_*=2\sqrt{3}$. If we put $x=x_*(1+\varepsilon\tilde{\zeta})$,
we obtain
\beqa
W(x)&=&-\frac{\sqrt{2}}{135}\varepsilon^{\frac{5}{2}}x_*^{5}
\chi_a^{\frac{5}{2}}\mu^{\frac{5}{4}}
(\Omega\mu^{-\frac{1}{2}}\tilde{\zeta}-\xi_{1})
(\Omega\mu^{-\frac{1}{2}}\tilde{\zeta}-\xi_{2})
\sqrt{\Omega\mu^{-\frac{1}{2}}\tilde{\zeta}+1} \n
&=&-\frac{\sqrt{2}}{540}
\varepsilon^{\frac{5}{2}}x_*^{5}\chi_a^{\frac{5}{2}}\mu^{\frac{5}{4}}
\sqrt{2}\eta_3(\Omega\mu^{-\frac{1}{2}}\tilde{\zeta})
\label{k=3W(x)}
\eeqa
where
\beqa
&&\chi_a=\frac{\sqrt{3}}{2}\sqrt{\beta}=\frac{1}{4}\alpha, \n
&&\Omega=\frac{1}{\chi_a}, \n
&&\xi_{1}=\frac{1-\sqrt{5}}{4}=-\cos\frac{2\pi}{5}, \n
&&\xi_{2}=\frac{1+\sqrt{5}}{4}=-\cos\frac{4\pi}{5}.
\eeqa
By integrating $W(x)$ over $x$, we obtain
\beqa
NV_{eff}^{(0)}(x)=-2N\int^x dx' \: W(x')
&=&\frac{\sqrt{2}}{270}g_s^{-1}\kappa^{-1}x_*^{6}\chi_a^{\frac{5}{2}}
\Omega^{-1}\mu^{\frac{7}{4}}v_3(\Omega\mu^{-\frac{1}{2}}\tilde{\zeta}) \n
&=&\frac{4\cdot 3^{\frac{7}{4}}}{5}g_s^{-1}\kappa^{-1}\beta^{\frac{7}{4}}
\mu^{\frac{7}{4}}v_3(\Omega\mu^{-\frac{1}{2}}\tilde{\zeta}).
\label{k=3NVeff0}
\eeqa
It is easy to verify that the last line in (\ref{k=3NVeff0})
can also be obtained by plugging $k=3$ and 
(\ref{alphaandnu}) into (\ref{NVeff0even}).

%%%%%%%%%%%%%%%%%%%%%%%%%%%%%%%%%%%%%%%%%%%%%%%%%%%%%%%%%%%%%%%%%
\section*{Appendix B: Useful formulae}
%%%%%%%%%%%%%%%%%%%%%%%%%%%%%%%%%%%%%%%%%%%%%%%%%%%%%%%%%%%%%%%%%
\setcounter{equation}{0}
\renewcommand{\theequation}{B.\arabic{equation}}
In this appendix, we gather some formulae, 
which we use in sections 5 and 6.

The following formulae for the Legendre polynomials are used in section 5.
\beqa
P_n(x)=\frac{1}{2^n}\sum_{p=0}^{[\frac{n}{2}]}
\frac{(-1)^p(2n-2p)!}{(n-2p)!p!(n-p)!}x^{n-2p}.
\label{Pn}
\eeqa
\beqa
P_n(-x)=(-1)^nP_n(x).
\label{Pn-}
\eeqa
\beqa
\int_{-1}^x dt\:(x-t)^{-\frac{1}{2}}P_n(t)
=\frac{1}{n+\frac{1}{2}}\frac{T_n(x)+T_{n+1}(x)}{\sqrt{x+1}}.
\label{formula}
\eeqa
\beqa
\int_0^1dx\: x^m P_n(x)
=\frac{m(m-1)\cdots(m-n+2)}{(m+n+1)(m+n-1)\cdots(m-n+3)}.
\label{formula2}
\eeqa

The cylinder contribution to the two macroscopic loop correlators
in the one matrix model with a general potential is obtained in
\cite{AJM}. The result is 
\beqa
\left\langle\tr\left(\frac{1}{x-\phi}\right)\tr\left(\frac{1}{y-\phi}\right)
\right\rangle_c
=\frac{1}{2(x-y)^2}
\left(\frac{xy-\frac{1}{2}(a+b)(x+y)+ab}
{\sqrt{(x-a)(x-b)(y-a)(y-b)}}
-1\right).
\eeqa
By integrating this over $x$ and $y$, we obtain
\beqa
&&\langle\tr\log(x-\phi)\:\tr\log(y-\phi)\rangle_c \n
&&=\int_x^{\infty}dx'\int_{y}^{\infty}dy'\:
\left\langle\tr\left(\frac{1}{x'-\phi}\right)\tr\left(\frac{1}{y'-\phi}\right)
\right\rangle_c \n
&&=\frac{1}{2}\log\left|2xy-(a+b)(x+y)+a^2+b^2
-2\sqrt{(x-a)(x-b)(y-a)(y-b)}\right| \n
&&\;\;\;\;-\log\left(\sqrt{(x-a)(x-b)}+\sqrt{(y-a)(y-b)}
\right) \n
&&\;\;\;\;+\log\left(\sqrt{x-a}+\sqrt{x-b}\right)
+\log\left(\sqrt{y-a}+\sqrt{y-b}\right)
-\log(a-b)-\log 2.
\label{trlntrln}
\eeqa

\end{document}